\documentclass[conference,letterpaper]{IEEEtran}
\pdfoutput=1

\usepackage[hyphens]{url}
\usepackage{graphicx}
\usepackage{paralist}
\usepackage{times}
\usepackage{xspace}
\usepackage{datetime}
\usepackage{soul}
\usepackage{cite}
\usepackage{color}

\long\def\com#1{}

\long\def\xxx#1{}

\defaultleftmargin{\parindent}{}{}{}

\newcommand{\eg}{{\em e.g.}}
\newcommand{\ie}{{\em i.e.}}

\newcommand{\coty}{cothority\xspace}
\newcommand{\coties}{cothorities\xspace}
\newcommand{\cosi}{CoSi\xspace}
\newcommand{\app}{CoSi\xspace}
\newcommand{\sigt}{Schnorr\xspace}

\newcommand{\hash}[1]{\mathrm{H}(#1)}

\begin{document}

\title{Keeping Authorities ``Honest or Bust'' \\
	with Decentralized Witness Cosigning}

\author{
\IEEEauthorblockN{Ewa Syta, Iulia Tamas, \\ Dylan Visher, David Isaac Wolinsky}
\IEEEauthorblockA{Yale University \\ New Haven, CT, USA}
\and
\IEEEauthorblockN{Philipp Jovanovic, Linus Gasser, \\
		Nicolas Gailly, Ismail Khoffi, Bryan Ford}
\IEEEauthorblockA{Swiss Federal Institute of Technology (EPFL) \\
			Lausanne, Switzerland}
}

\maketitle

\begin{abstract}

The secret keys of critical network authorities --
such as time, name, certificate, and software update services --
represent high-value targets for hackers, criminals, and spy agencies
wishing to use these keys secretly to compromise other hosts. 
To protect authorities and their clients proactively
from undetected exploits and misuse,
we introduce \cosi, 
a scalable {\em witness cosigning} protocol
ensuring that every authoritative statement
is validated and publicly logged by a diverse group of witnesses
before any client will accept it.
A statement $S$ collectively signed by $W$ witnesses
assures clients that $S$ has been seen,
and not immediately found erroneous,
by those $W$ observers.
Even if $S$ is compromised in a fashion
not readily detectable by the witnesses,
\cosi still guarantees $S$'s exposure to public scrutiny,
forcing secrecy-minded attackers to risk
that the compromise will soon be detected by one of the $W$ witnesses.
Because clients can verify collective signatures efficiently
without communication,
\cosi protects clients' privacy, and
offers the first transparency mechanism effective against
persistent man-in-the-middle attackers who control
a victim's Internet access, the authority's secret key,
and several witnesses' secret keys.
\cosi builds on existing cryptographic multisignature methods,
scaling them to support thousands of witnesses
via signature aggregation over efficient communication trees.
A working prototype demonstrates \cosi
in the context of timestamping and logging authorities,
enabling groups of over 8,000 distributed witnesses
to cosign authoritative statements in under two seconds.

\end{abstract}

\section{Introduction}

Centralized {\em authorities} provide
critical services that many hosts and users rely on,
such as time~\cite{mills91internet}
and timestamp services~\cite{adams01internet},
certificate authorities (CAs)~\cite{chokhani99internet},
directory authorities~\cite{sermersheim06lightweight,dingledine04tor},
software update services~\cite{samuel10survivable},
digital notaries~\cite{adleman83implementing},
and randomness services~\cite{beacon,random-org}.
Even when cryptographically authenticated,
authorities represent central points of failure
and attractive attack targets for hackers, criminals, and spy agencies.
Attackers obtaining the secret keys of any of
hundreds of CAs~\cite{eff11observatory}
can and have misused CA authority
to impersonate web sites and spy on 
users~\cite{bright11diginotar,bright11comodo,arthur11diginotar,tung15google}.
By impersonating a time service an attacker can
trick clients into accepting expired certificates
or other stale credentials~\cite{malhotra15attacking}.
Criminals increasingly use stolen code-signing keys
to make their malware appear trustworthy~\cite{kessem15certificates}.

Logging and monitoring proposals such as
Perspectives~\cite{wendlandt08perspectives},
CT~\cite{rfc6962, laurie14CT},
AKI~\cite{kim13accountable},
ARPKI~\cite{basin14arpki}, and
PoliCert~\cite{szalachowski14policert}
enable clients to cross-check certificates against public logs,
but this checking requires active communication.
To avoid delaying web page loads
this checking is usually done only retroactively,
leaving a time window an attacker could exploit to serve the client
malware or backdoored software,
which can then disable detection.
An attacker who controls the client's access network --
such as a compromised home or corporate network,
or an ISP controlled by authoritarian state --
can block access to uncompromised log servers,
permanently evading detection
if the targeted client is not sufficiently mobile.
Finally, checking logs can create privacy concerns
for clients~\cite{melara15coniks,nordberg15gossiping},
and the log servers themselves become new
central points of failure that must be audited~\cite{nordberg15gossiping}.

\com{
The key observation is that public authorities are normally intended to
sign {\em public} statements
such as the current time,
a public key representing a domain name,
or the latest version of a software package --
whereas powerful attackers who might steal an authority's secret key
usually wish to use the key {\em in secret} without detection,
to perform man-in-the-middle (MITM) attacks on targeted clients for example.
}

To address these weaknesses we propose {\em witness cosigning},
a proactive approach to transparency
that can either replace or complement existing approaches.
When an authority publishes a new signing key,
to be bundled in a web browser's set of root certificates for example,
the authority includes with it the identities and public keys of
a preferably large, diverse, and decentralized group
of independent {\em witnesses}.
Whenever the authority subsequently signs a new authoritative statement
such as a new timestamp, certificate, or log record,
the authority first sends the proposed statement to its witnesses
and collects {\em cosignatures},
which the authority attaches to the statement together with its own signature.
A client receiving the statement (\eg, as a TLS certificate)
verifies that it has been signed not only by the authority itself
but also by an appropriate subset of the witnesses.
The client's signature acceptance criteria may be
a simple numeric threshold (\eg, 50\% of the witnesses)
or a more complex predicate accounting for trust weights,
groupings of witnesses, or even contextual information
such as whether a signed software update
is to be installed automatically or by the user's explicit request.

Witness cosigning offers clients direct cryptographic evidence --
which the client can check efficiently without communication --
that many independent parties
have had the opportunity to validate and publicly log
any authoritative statement
before the client accepts it.
Without witness cosigning, an attacker who knows the authority's secret key
can use it in man-in-the-middle (MITM) attacks against targeted victims,
anywhere in the world and without the knowledge of the legitimate authority,
to feed the victim faked authoritative statements
such as TLS certificates or software updates\cite{ryge16most}.
To attack a client who demands that statements
be cosigned by at least $W$ witnesses, however,
a MITM attacker must either
(a) control both the authority's secret key and those of $W$ witnesses,
which becomes implausible if $W$ is sufficiently large and diverse, or
(b) submit the faked statement to one or more honest witnesses for cosigning,
thereby exposing the faked statement to public scrutiny and risking detection.

We do not expect witnesses to detect all malicious statements immediately:
for example, only a CA itself may have the information needed to verify
the true correspondence between a name and a public key.
Witnesses can, however, sanity-check the correctness and consistency
of proposed statements before cosigning:
\eg, that authoritative timestamps are not wildly different
from the witnesses' view of the present time,
that logging authorities sign records in sequence
without revising history or equivocating~\cite{levin09trinc},
or that only one authoritative binary image exists
for a given software version number.
Even if witnesses cannot immediately tell
which of two conflicting TLS certificates or binaries is ``good,''
they can ensure that the existence of the conflicting signed statements
promptly becomes public knowledge.
Witnesses can proactively check that statements conform to known policies,
such as certificate issuance policies~\cite{szalachowski14policert},
raising alarms and withholding their cosignature if not.
Finally, witnesses can of course publish logs
of statements they cosigned to increase 
the likelihood of rapid attack detection~\cite{laurie14CT}.

Even if witnesses perform little or no validation of the authority's statements,
their proactive presence in statement signing deters attackers
both by increasing the {\em threat} to the attacker of rapid misuse detection,
and by reducing the effective {\em value} of an authority's secret keys
to attackers wishing to operate in secret.
Witness cosigning thus serves as a ``Ulysses pact''
between the authority and its witnesses~\cite{doctorow16using}.

Authorities could implement witness cosigning
simply by collecting and concatenating individual signatures from witnesses,
exactly like PGP~\cite{rfc4880} or Bitcoin~\cite{nakamoto08bitcoin}
can already attach multiple signatures to a message or transaction.
This is practical with
tens or perhaps even a few hundred witnesses, 
but incurs substantial signature size and verification costs
as the witness group grows large.
To make witness cosigning scalable we introduce \cosi, 
a witness cosigning protocol enabling authoritative statements
to be validated and cosigned by thousands of witnesses in a few seconds,
to produce collective signatures comparable in size to
a single individual signature (\eg, $\approx 100$ bytes total)
and nearly as quick and easy for clients to verify.

As a scenario motivating \cosi's scalability goal,
we envision the DNS\-SEC~\cite{rfc4033} root zone
might be witnessed by all willing operators of
the now over 1,000 top-level domains (TLDs).
Future TLS certificates might be witnessed
by all other willing CAs, of which there are hundreds~\cite{eff11observatory},
and by other parties such as
CT servers~\cite{laurie14CT}.
Public ledgers of
national cryptocurrencies~\cite{morris14inside,sputnik15russia}
might be collectively witnessed by all willing banks in the country --
of which the US has thousands
even after consolidation~\cite{tracy13tally}.
Threshold signatures~\cite{micali01accountable,bellare06multi}
and consensus protocols~\cite{castro99practical,tor-directory}
can split trust across a few nodes (typically 3--10),
but do not scale, as we confirm in Section~\ref{sec:evals}.
To our knowledge \cosi is the first multisignature protocol
that scales to thousands of signers.

\cosi's scalability goal presents 
three key technical challenges:
efficient cosignature collection,
availability in the face of slow or offline witnesses,
and efficient cosignature verification by clients.
\cosi makes verification efficient by adapting
well-understood Schnorr multisignatures~\cite{okamoto99multi}
to combine many cosignatures into a single compact signature,
typically less than $100$ bytes in size,
which clients can check in constant time.
To collect and combine thousands of cosignatures efficiently,
\cosi adapts tree-based techniques, long used in
multicast~\cite{deering90multicast,castro03splitstream,venkataraman06chunkyspread},
and aggregation protocols~\cite{cappos08sanfermin,yalagandula04sdims}
to scalable multisignatures.
To protect the authority's availability
even when witnesses go offline,
\cosi includes metadata in its collective signatures
to document ``missing witnesses''
and enable verifiers to check the signature correctly
against an aggregate of the remaining witnesses' public keys.


\com{
The formal foundation for cothorities
already exists in the form of cryptographic
threshold signatures~\cite{shoup00practical},
aggregate signatures~\cite{boneh03survey}, and 
multisignatures~\cite{micali01accountable,bellare06multi},
but to our knowledge these primitives have been deployed only in small groups
(\eg, $\approx 10$ nodes) in practice.
\xxx{	Review A: can we offer citations? }
Our main contribution is to demonstrate how to scale these techniques
across thousands of servers in practical environments.
A first-order technical challenge is limiting the
computation and network bandwidth costs imposed on each participating server;
we solve this challenge using tree-based communication structures
comparable to those long used in
multicast protocols~\cite{castro03splitstream,venkataraman06chunkyspread}.

A second-order challenge is handling server failures,
which we expect to be rare but non-negligible,
and if not addressed would make a cothority vulnerable to
crashes or denial-of-service attacks by any server.
We explore two solutions to this availability challenge.
First, we allow each log entry's signature to contain a few {\em exceptions}
explicitly listing servers whose contributions to the collective signature
could not be obtained promptly.
Second, we can require cothority servers to secret-share their
temporary signing keys with a group of independent {\em insurers},
who can reconstruct the server's key if the server fails.
These two approaches to guaranteeing availability represent tradeoffs
and may be employed independently or together.
}

We have built a working \cosi prototype,
deployed a small-scale test configuration on the public Internet,
and evaluated it at larger scales of up to 33,000 cosigning witnesses
on the DeterLab~\cite{deterlab} testbed.
We find that \cosi can collect and aggregate cosignatures
from 8,000 witnesses,
separated by 200ms round-trip network delays to simulate distribution,
in about 2 seconds total per signing round.
\cosi's performance contrasts favorably with
multisignatures produced via classic
verifiable secret sharing (VSS)~\cite{feldman87practical,stadler96publicly},
whose signing costs explode beyond about 16 participants,
as well as with straightforward collection of individual cosignatures,
whose costs become prohibitive beyond around 256 participants.

In addition, we have integrated \cosi into
and evaluated it in the context of two specific types of ``authorities'':
a secure time and timestamping service~\cite{haber91how,adams01internet,sirer13introducing},
and the Certificate Transparency log server~\cite{laurie14CT}.
The \cosi timestamping service illustrates how some authorities
can be made even more scalable by building on \cosi's communication trees,
allowing witnesses to serve timestamp requests
and reduce load on the main authority,
thereby achieving aggregate throughput
of over 120,000 timestamp requests per second in a 4,000-witness configuration.
The \cosi extension to the CT log server demonstrates
the ease and simplicity with which witness cosigning
can be added to existing authority services,
in this case requiring only an 315-line change to the log server
to invoke \cosi when signing each new log entry.

\com{
We further evaluate this cothority server prototype informally in the context
of three realistic applications:
(a) to create a Certificate Cothority (CC) through which the hundreds of
present-day CAs could in principle monitor and collectively sign
each others' certificates to create a {\em proactively trusted}
public log of certificates;
(b) to create a more secure alternative
to the NIST Randomness Beacon~\cite{beacon};
and 
(c) to implement a large-scale, Byzantine-fault-tolerant directory service
of the type needed to create a more secure Tor directory authority~\cite{torDPv3}.
}

In summary, this paper contributes:
(a) a proactive approach to transparency based on witness cosigning;
(b) \cosi, the first collective signing protocol that 
demonstrably scales to thousands of participants;
(c) an experimental implementation of \cosi that demonstrates its practicality
and how it can be integrated into existing authority services.

Section~\ref{sec:bg} of this paper explores the background 
and motivation for witness cosigning. 
Section~\ref{sec:sign} then presents \app,
a scalable collective signing protocol.
Section~\ref{sec:var} outlines
variants of the \cosi design offering different tradeoffs.
Section~\ref{sec:impl} describes the details of our 
prototype implementation of \cosi
and its incorporation into timestamping and certificate logging applications.
Section~\ref{sec:evals} experimentally evaluates this prototype, and
Section~\ref{sec:disc} discusses \cosi's applicability
to real-world applications and outlines future work.
Section~\ref{sec:rel} summarizes related work and 
Section~\ref{sec:conc} concludes.


\section{Background and Motivation}
\label{sec:bg}

This section briefly reviews several types of conventional authorities,
their weaknesses,
and how witness cosigning can help strengthen them.
We revisit prototype implementations of some of these applications
later in Section~\ref{sec:impl}.

\com{	redundant with tamper-evident logging text.
Authorities may misbehave in an arbitrary way, however, 
the main threats include 
(a) rewriting history, when the authority rolls back pretending
that a different decisions was made or that a decision was never made;
(b) equivocation, when the authority gives different clients different views of history;
(c) affecting the future, when the authority deviates from the specification
when making its decisions.
}

\subsection{Certificate Authorities and Public-Key Infrastructure}

\com{ from Cappos: Okay, a minor suggestion: You might clarify that it can be used either way (or pick a way) for the paper.  It definitely threw me off because I was expecting you had one or the other in mind and I could not understand which.}

Certificate Authorities (CAs)
sign certificates attesting that a public key
represents a name such as \verb|google.com|,
to authenticate SSL/TLS connections~\cite{rfc5246,rfc6101}.
Current web browsers directly trust dozens of root CAs
and indirectly trust hundreds of
intermediate CAs~\cite{eff11observatory},
any one of which can issue fake certificates
for any domain if compromised.
\com{	superfish is more about a bad software install than CA structure...
Additionally, there is no single list of
``approved CAs"; each browser or operating system can introduce 
their own root certificates, an issue abundantly illustrated by 
the Lenovo's Superfish snafu~\cite{rosenblatt15superfish}.
}
Due to this ``weakest-link'' security,
hackers have stolen the ``master keys'' of CAs such as
DigiNotar~\cite{bright11diginotar,arthur11diginotar} and
Comodo~\cite{bright11comodo}
and abused certificate-issuance
mechanisms~\cite{langley13anssi,langley15cnnic,tung15google}
to impersonate popular websites and attack their users.

\com{
Related but weaker motivation:
"Analyzing Forged SSL Certificates in the Wild"
measured forged certs, most from anti-virus software and corporate filters,
but some intercepted by malware.
}

\com{
Multipath approaches:
Perspectives~\cite{wendlandt08perspectives},
Convergence~\cite{marlinspike11ssl,bates14forced},

Logging approaches:
CT~\cite{rfc6962, laurie14CT},
AKI~\cite{kim13accountable},
ARPKI~\cite{basin14arpki},
PoliCert~\cite{szalachowski14policert}.

DoubleCheck~\cite{alicherry09doublecheck}

generalized pinning:
Certlock~\cite{soghoian11certified},
TACK~\cite{marlinspike13trust}.
DVCert~\cite{dacosta12trust},

}

As a stopgap,
some browsers hard-code or
{\em pin} public keys for popular sites
such as \verb|google.com|~\cite{rfc7469} --
but browsers cannot hard-code public keys for the whole Web.
Related approaches offer TOFU (``trust on first use") security
by pinning the first public key a client
sees for a particular site~\cite{soghoian11certified,dacosta12trust,marlinspike13trust},
thereby protecting regular users but not new users.
Browsers can check server certificates against
public logs~\cite{rfc6962,laurie14CT,kim13accountable,
		basin14arpki,szalachowski14policert,ryan14enhanced},
which independent monitors may check for invalid certificates.
Monitoring can unfortunately detect misbehavior only retroactively,
placing victims in a race with the attacker.
Browsers could check certificates
against such logs and/or via multiple Internet
paths~\cite{wendlandt08perspectives,marlinspike11ssl,
		bates14forced,alicherry09doublecheck},
but such checks delay the critical page-loading path,
at least on the first visit to a site.
Further,
these approaches assume Web users can connect to
independent logging, monitoring, or relaying services without interference,
an assumption that fails when the user's own ISP is compromised.
Such scenarios are unfortunately all too realistic
and have already occurred,
motivated by state-level
repression~\cite{bright11diginotar,arthur11diginotar}
or commercial interests~\cite{hoffman-andrews14verizon,feinberg15gogo}.

A CA might arrange for a group of witnesses
to cosign certificates it issues:
\eg, other willing CAs and/or independent organizations.
Witness cosigning might not only proactively protect users
and increase the CA's perceived trustworthiness,
but also decrease the value of the CA's secret keys to potential attackers
by ensuring that any key misuse is likely to be detected quickly.
In the longer term,
CAs might witness cosign OCSP staples~\cite{rfc6961},
or entire key directory snapshots
as in CONIKS~\cite{melara15coniks},
enabling clients to check not only the validity
but also the freshness of certificates
and address persistent weaknesses in
certificate revocation~\cite{liu15end}.

\com{
By federating today's hundreds of CAs
into a single {\em certificate cothority},
each CA could in principle validate certificates
proposed by all other CAs {\em before} they are collectively signed.
For example, the CA currently responsible for a given domain
such as \verb|google.com|
could verify that no other CA proposes a \verb|google.com| certificate,
raising an alarm and proactively preventing the signing
of fake certificates in the first place.
Each certificate would still be validated by a single digital signature,
but that signature would embody much stronger and broader-based {\em trust}
of the entire certificate cothority. 
The goal of this approach is to bring transparency to the action's performed by
each individual certificate authority by making all certificates publicly witnessed
by the cothority members. In this scenario, an attacker cannot create a rogue certificate 
that has not been widely witnessed. 
It is not a goal, however, to shift the individual responsibility 
of each CA in charge of a particular domain to do the proper validity checking 
associated with different types of certificates that might be issued. 
}

\subsection{Tamper-Evident Logging Authorities}

Many storage systems and other services rely on tamper-evident
logging~\cite{li04sundr,crosby09efficient}. 
Logging services are vulnerable to {\em equivocation}, however,
where a malicious log server
rewrites history or
presents different ``views of history'' to different clients.
Even if a logging authority itself is well-behaved,
an attacker who obtains the log server's secret keys
can present false logs to targeted clients,
effectively ``equivocating in secret''
without the knowledge of the log's legitimate operator.
For example, an attacker can defeat CT~\cite{laurie14CT}
and attack clients this way
by secretly stealing the keys of -- or coercing signatures from --
any single CA plus any two CT log servers.

Solutions to equivocation attacks
include weakening consistency guarantees
as in SUNDR~\cite{li04sundr},
or adding trusted hardware
as in TrInc~\cite{levin09trinc}.
Equivocation is the fundamental reason
Byzantine agreement in general requires $N=3f+1$ total nodes
to tolerate $f$ arbitrary failures~\cite{castro99practical}.
Witness cosigning does not change this basic situation,
but can make it practical for both $N$ and $f$ to be large:
\eg, with $N>3000$ participants independently checking
and cosigning each new log entry,
arbitrarily colluding groups up to $1000$ participants
cannot successfully equivocate or rewrite history.
As a proof-of-concept,
Section~\ref{sec:impl:log} later presents
such a witness cosigning extension
for Certificate Transparency log servers.

\subsection{Time and Timestamping Authorities}

Time services such as NTP~\cite{mills91internet,rfc5905}
enable hosts to learn the current time and synchronize their clocks
against authoritative sources such as NIST's
Internet Time Service~\cite{lombardi02nist}.
Cryptographic authentication was a late addition to NTP~\cite{rfc5906}
and is still in limited use,
leading to many vulnerabilities~\cite{malhotra15attacking}.
For example,
an attacker impersonating a legitimate time service
might falsify the current time,
to trick a client into accepting an expired certificate
or other stale credentials.

A timestamping authority~\cite{haber91how,adams01internet}
enables a client to submit
a cryptographic hash or commitment to some document
(\eg, a design to be patented),
and replies with a signed statement attesting
that the document commitment was submitted at a particular date and time.
The client can later prove to a third-party that the document existed
at a historical date by opening the cryptographic commitment
and exhibiting the authority's timestamped signature on it.
Virtual Notary~\cite{sirer13introducing} generalizes timestamp services
by offering users timestamped attestations
of automatically checkable online facts
such as web page contents, stock prices, exchange rates, etc.
An attacker who steals a timestamp service's secret keys
can forge pre-dated timestamps on any document, however,
and a notary's secret key similarly enables an attacker
to create legitimate-looking attestations of fake ``facts.''

While witness cosigning incurs communication latencies
that likely preclude its use in fine-grained clock synchronization,
it can serve a complementary role of increasing the security
of {\em coarse-grained} timestamps,
\ie,
giving clients greater certainty
that a timestamp is not hours, days, or years off.
Section~\ref{sec:impl:time} later presents a prototype of such a service,
in which many witnesses efficiently sanity-check batches of signed timestamps,
ensuring that even an attacker who compromises the authority's secret key
cannot undetectably back-date a timestamp beyond a limited time window.

\subsection{Directory Authorities}

The Domain Name System (DNS)~\cite{mockapetris88development,rfc1034}
offers a critical directory service for locating Internet hosts by name.
Like NTP,
DNS initially included no cryptographic security;
even now the deployment of DNSSEC~\cite{rfc4033} is limited
and weaknesses remain~\cite{ariyapperuma07security}.
The fact that DNSSEC is completely dependent
on the security of its Root Zone~\cite{rfc3833},
which is centrally managed by one organization,
is a concern despite measures
taken to secure the Root Zone's signing keys~\cite{rfc6781}.
If Root Zone signatures were witnessed and cosigned by
all willing operators of subsidiary top-level domains (TLDs),
ensuring rapid discovery of any misuse of the Root Zone's keys,
concerns about DNSSEC's centralization might be alleviated.

As another example,
clients of the Tor anonymity system~\cite{torproject}
rely on a directory authority~\cite{tor-directory}
to obtain a list of available anonymizing relays.
A compromised Tor directory authority could give clients
a list containing only attacker-controlled relays, however,
thereby de-anonymizing all clients.
To mitigate this risk, Tor clients accept a list
only if it is signed by a majority of a small {\em consensus group},
currently nine servers.
Because these directory servers and their private directory-signing keys
represent high-value targets for
increasingly powerful
state-level adversaries~\cite{khrennikov14putin,greenberg15tor},
it is questionable whether
a small, relatively centralized group offers adequate security.
If Tor directory snapshots were witness cosigned
by a larger subset of the thousands of regular Tor relays,
the risk of semi-centralized directory servers being silently compromised
might be reduced.

\subsection{Software Download and Update Authorities}
\label{sec:bg:update}

App stores, community repositories, and
automatic software update services
have become essential in patching security vulnerabilities promptly.
Update services themselves can be
attack vectors,
however~\cite{bellissimo06secure,cappos08look,nullbyte14hack,ryge16most}.
Even when updates are authenticated,
code signing certificates are
available on the black market~\cite{kessem15certificates},
and software vendors have even leaked their secret keys
accidentally~\cite{mimoso15dlink}.
Governments desiring backdoor access
to personal devices~\cite{ackerman15fbi,burgess15un},
as well as resourceful criminals,
might coerce or bribe vendors to sign
and send compromised updates to particular users.
These risks are exacerbated by the fact that
automatic update requests can amount to public announcements
that the requesting host is unpatched,
and hence vulnerable~\cite{cappos13avoiding}.
By witness cosigning their updates
and checking cosignatures in auto-update mechanisms,
software vendors might alleviate such risks
and ensure the prompt detection of any improperly signed software update.

\subsection{Public Randomness Authorities}

Randomness authorities~\cite{beacon,random-org}
generate non-secret random numbers or coin-flips,
which are useful for many purposes such as lotteries, sampling,
or choosing elliptic curve parameters~\cite{lenstra15random}.
NIST's Randomness Beacon~\cite{beacon}, for example,
produces a log of signed, timestamped random values from a hardware source.
If compromised, however, a randomness authority
could deliberately choose its ``random'' values as to win a lottery,
or could look into the future
to predict a lottery's outcome~\cite{weissman15how}.
In the wake of the DUAL-EC-DRBG debacle~\cite{checkoway14practical},
the NIST beacon has been skeptically labeled
``the \st{NSA}NIST Randomness Beacon''~\cite{beacon14se} and
``Project {`Not a backdoor'}''~\cite{beacon14reddit}.
While witness cosigning alone would not
eliminate all possibility of bias~\cite{lenstra15random,bonneau15bitcoin},
witnesses could preclude randomness beacons from revising history --
and by mixing entropy provided by witnesses into the result,
witnesses can ensure that even a compromised beacon
cannot predict or exercise unrestricted control over future ``random'' outputs.

\section{Scalable Collective Signing}
\label{sec:sign}

This section presents \cosi,
the first collective signing protocol
efficiently supporting large-scale groups.
We first outline \cosi's high-level principles of operation,
then detail its design,
covering a number of challenges such as
unavailable witnesses, cothority certificate size,
denial-of-service (DoS) risks and mitigations,
and statement validation by witnesses.

\subsection{Architecture and Principles of Operation}
\label{sec:sign:arch}

\begin{figure}[t]
\center\includegraphics[width=0.40\textwidth]{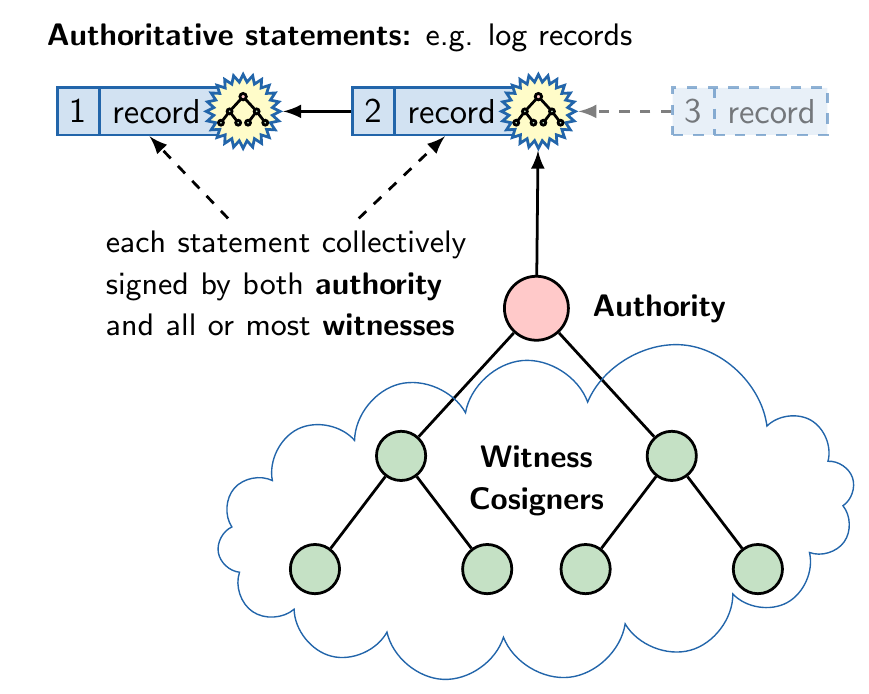}
\caption{\cosi protocol architecture.}
\label{fig:arch}
\end{figure}

Figure~\ref{fig:arch} illustrates
\cosi's conceptual architecture,
consisting of an authority who regularly signs statements of any kind
(\eg, chained log records in the example shown),
and a group of {\em witness cosigners} 
who participate in the signing of each record.
We also refer to the group of witnesses as a {\em witness cothority}:
a ``collective authority'' whose purpose is to
witness, validate, and then cosign the authority's statements.

The authority serves as the \cosi protocol's {\em leader},
defining and publishing the witness \coty's composition,
initiating collective signing rounds,
and proposing statements to be signed such as
timestamps, directories, or certificates.
We assume the witnesses to be
reliable, in\-depen\-dent\-ly-run servers maintained
by individuals or organizations who have agreed
to witness the leader's authoritative statements.
Realistic authorities typically serve clients as well:
\eg, users requesting timestamps or certificates.
In the basic \cosi architecture these clients interact
only with the authority (leader)
so we will ignore them for now,
although Section~\ref{sec:impl:time} will illustrate
how some types of authorities can leverage \cosi
to distribute client servicing load across the many witnesses.

We assume that the authority's group of witnesses is fixed or changes slowly,
and that all participants including cosignature verifiers
know both the authority's and all witnesses' public keys.
If the authority is a root CA
that signs TLS certificates to be verified by web browsers,
for example,
then the CA's root certificate shipped with the browser
includes a list of the public keys of the witnesses
in addition to the CA's own public key.
We assume the authority arranges for the witness list to remain valid
for a significant time period --
\eg, three years or more, comparable to root certificate lifetimes --
and that software updates can handle witness list evolution
just as for root certificates.
If the size of the authority's root certificate
and its witness list becomes an issue,
it may be compressed into a cryptographic hash of that roster,
at a cost of increased signature sizes as discussed later
in Section~\ref{sec:live:rep}.
For security reasons discussed later in Section~\ref{sec:schnorr}
we require that the public keys of the authority and all witnesses
be self-signed to prove knowledge of the corresponding secret key.

\subsection{Threat Model}
\label{sec:sign:threat}

We assume both the authority (leader)
and some number of the authority's witnesses may be malicious and colluding
in attempts to sign malicious statements secretly
that unsuspecting victims (verifiers) will accept,
without these malicious statements being detected by honest witnesses.
The \cosi protocol does not assume or specify
any particular global cosignature verification threshold,
but from the perspective of a client
who demands at least $f+1$ cosignatures on a statement,
we assume the attacker controls at most $f$ faulty witnesses.

We assume the authority (leader) is live and highly available:
since it is the participant who wishes to produce witnessed statements,
\cosi makes no attempt to protect against DoS by the leader.
However, we assume that a threshold number of witnesses may go offline
at any time or even engage in DoS attacks;
this threshold is a policy parameter defined by the leader.
Witnesses may also maliciously produce incorrect messages
deviating from the protocol,
\eg, in attempt to trick the leader into misbehavior.
While for now we assume simple numeric thresholds,
clients can impose more complex verification predicates if desired
(Section~\ref{sec:sig-verify}).

We assume the leader and all witnesses
are generally able to communicate with each other,
apart from temporary communication outages.
Unlike gossip-based transparency approaches,
however,
we do {\em not} assume that clients verifying signatures
can communicate with any non-attacker-controlled parties.

\subsection{Responsibilities of Cosigning Witnesses}
\label{sec:sign:witnesses}

The authority determines when to initiate a collective signing round,
and broadcasts to all witnesses the statement to be signed.
Witnesses may, and ideally should,
publish logs of the statements they witness and cosign,
thus serving a transparency role
similar to log servers in CT~\cite{rfc6962, laurie14CT}.
If the authority's statements are already supposed to take the form of a log
as in the example in Figure~\ref{fig:arch},
then each witness might simply make available a public mirror
of all or some recent portion of the authority-generated log.

Witnesses may also, and ideally should,
perform any readily feasible syntactic and semantic correctness checks
on the authority's proposed statements
before ``signing off'' on them.
If the authority's statements include a wall-clock timestamp, for example,
each witness may verify that the proposed timestamp
is not wildly different from the witness's view of the current time
(\eg, is not minutes or hours off).
If the authority's statements form a sequence-numbered, hash-chained log
as in Figure~\ref{fig:arch},
each witness may verify that each of the authority's proposed log records
contains a monotonically increasing sequence number
and the correct hash for the immediately preceding log record,
preventing a compromised authority from reversing or rewriting history.\footnote{
Even with these checks a faulty authority could still {\em equivocate}
to produce two or more divergent histories
cosigned by disjoint subsets of honest witnesses.
Applying standard Byzantine consensus principles~\cite{castro99practical},
however,
the above log consistency checks will preclude equivocation provided
at most $f$ witnesses are faulty out of at least $3f+1$ total,
and provided verifiers check that
at least $2f+1$ witnesses have cosigned each statement.}

Witnesses might check deeper application-specific invariants as well,
provided these checks are quick and automatic.
If the authority's statements represent certificates,
witnesses may check them against any known issuance policies
for the relevant domain~\cite{szalachowski14policert}.
If the authority's statements attest certificate freshness~\cite{rfc6961}
or represent directories of currently-valid certificates
as in CONIKS~\cite{melara15coniks},
witnesses may verify that these certificates
do not appear on cached certificate revocation lists (CRLs)~\cite{liu15end}.
If the authority's statements form a blockchain~\cite{nakamoto08bitcoin},
then witnesses may check its validity:
\eg, that each transaction is properly formed, properly authorized,
and spends only previously-unspent currency~\cite{kokoris16enhancing}.
If the authority's statements represent
software binaries~\cite{samuel10survivable},
then witnesses might even attempt to reproduce the proposed binaries
from developer-signed sources~\cite{bobbio14reproducible},
provided the authority allows the witnesses the time required
(possibly hours) to perform such builds during signing process.

For simplicity,
we initially assume that witnesses never fail or become disconnected,
but relax this unrealistic assumption
later in Section~\ref{sec:excep}.
We also defer until later performance concerns such as
minimizing collective signing latency.

\subsection{Schnorr Signatures and Multisignatures}
\label{sec:schnorr}

While \cosi could in principle build on many digital signature schemes
that support efficient public key and signature aggregation,
we focus here on one of the simplest and most well-understood schemes:
Schnorr signatures~\cite{schnorr90efficient}
and multisignatures~\cite{micali01accountable,bellare06multi}.
Many alternatives are possible:
\eg, Boneh-Lynn-Shacham (BLS)~\cite{boneh01short}
requires pairing-based curves,
but offers even shorter signatures (a single elliptic curve point),
and a simpler protocol that may be more suitable in extreme situations
as discussed later in Section~\ref{sec:async}.

Schnorr signatures rely on a group $\mathcal{G}$ of prime order $q$
in which the discrete logarithm problem is believed to be hard;
in practice we use standard elliptic curves for $\mathcal{G}$.
Given a well-known generator $G$ of $\mathcal{G}$,
each user chooses a random secret key $x < q$,
and computes her corresponding public key $X = G^x$.
We use multiplicative-group notation for consistency
with the literature on Schnorr signatures,
although additive-group notation may be more natural
with elliptic curves.

Schnorr signing is conceptually
a {\em prover-verifier} or $\Sigma$-protocol~\cite{damgard02sigma},
which we make non-interactive using the Fiat-Shamir heuristic~\cite{fiat87prove}.
To sign a statement $S$, the prover picks a random secret $v < q$,
computes a {\em commit}, $V = G^v$, and sends $V$ to the verifier.
The verifier responds with a random {\em challenge} $c < q$,
which in non-interactive operation is simply
a cryptographic hash $c = \hash{V \parallel S}$.
The prover finally produces a {\em response},
$r = v - cx$, where $x$ is the prover's secret key.
The challenge-response pair $(c,r)$ is the Schnorr signature,
which anyone may verify using the signer's public key $X = G^x$,
by recomputing $V' = G^r X^c$
and checking that $c \stackrel{?}{=} \hash{V' \parallel S}$.

With Schnorr multisignatures~\cite{okamoto99multi},
there are $N$ signers with individual secret keys $x_1,\dots,x_N$
and corresponding public keys $X_1=G^{x_1},\dots,X_N=G^{x_N}$.
We compute an {\em aggregate} public key $X$ from the individual public keys
as $X = \prod_i{X_i} = G^{\sum_i{x_i}}$.
The $N$ signers collectively sign a statement $S$ as follows.
Each signer $i$ picks a random secret $v_i < q$,
and computes a commit $V_i = G^{v_i}$.
One participant (\eg, a leader) collects all $N$ commits,
aggregates them into a collective commit $V = \prod_i{V_i}$,
and uses a hash function to compute a collective challenge
$c = \hash{V \parallel S}$.
The leader distributes $c$ to the $N$ signers,
each of whom computes and returns its response share
$r_i = v_i - cx_i$.
Finally, the leader aggregates the response shares into $r = \sum_i{r_i}$,
to form the collective signature $(c,r)$.
Anyone can verify this constant-size signature against the statement $S$
and the aggregate public key $X$
via the normal Schnorr signature verification algorithm.

When forming an aggregate public key $X$
from a roster of individual public keys $X_1,\dots,X_N$,
all participants must validate each individual public key $X_i$
by requiring its owner $i$ to prove knowledge
of the corresponding secret key $x_i$,
\eg, with a zero-knowledge proof or a self-signed certificate.
Otherwise, a dishonest node $i$ can perform a {\em related-key attack}~\cite{michels96risk}
against a victim node $j$ by choosing $X_i = G^{x_i} X_j^{-1}$,
and thereafter contribute to collective signatures apparently signed by $j$
without $j$'s actual participation.

While multisignatures are well-under\-stood
and formally analyzed,
to our knowledge they have so far been used or considered practical
only in small groups (\eg, $N \approx 10$).
The next sections describe how we can make multisignatures scale
to thousands of participants,
and address the availability challenges that naturally arise in such contexts.

\subsection{Tree-based Collective Signing}
\label{sec:tree}

\begin{figure*}[t]
\center\includegraphics[width=1.00\textwidth]{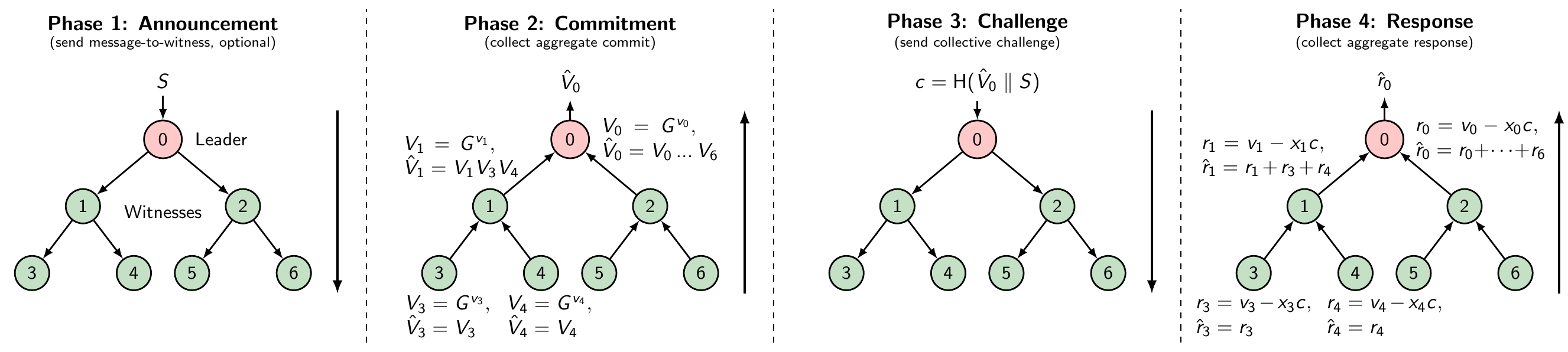}
\caption{The \cosi protocol uses four communication phases
	for scalable construction of a Schnorr multisignature $(c,\hat{r}_0)$
	over a spanning tree.}
\label{fig:phases}
\end{figure*}

To make multisignatures scale to many participants,
\cosi distributes the communication and computation costs of multisignatures
across a spanning tree analogous to those
long utilized in
multicast protocols~\cite{deering90multicast,castro03splitstream,venkataraman06chunkyspread}.
The leader organizes the $N$ witnesses 
into a spanning tree of depth $O(\log N)$ rooted at the leader,
distributing both communication and computation
to incur at most logarithmic costs per node.
The spanning tree serves only to optimize performance:
the leader may reconfigure it at any time without affecting security,
\eg, to account for unavailable witnesses
as detailed later in Section~\ref{sec:live}.

\com{Linus: for the moment we don't chose the tree-depth to be
O(logN)
	I think we do, at least from what I see in the graphs...
	big-O notation effectively means "order of",
	not some particular exact equation, so I think we're fine. -baf 
For the round-comparison we put all the protocols in a tree of depth
3 but varied the branching factor, getting between 3 and 33'000
witnesses...
}

For simplicity, the tree may be a regular $B$-ary tree
formed deterministically from the well-known list of $N$ witnesses,
thereby requiring no communication of the tree structure.
To minimize signing latency,
the leader might alternatively collect information
on round-trip latencies between witnesses,
construct a shortest-path spanning tree,
and specify this tree explicitly when announcing a collective signing round.

\com{
\begin{figure}[t]
\includegraphics[width=0.49\textwidth]{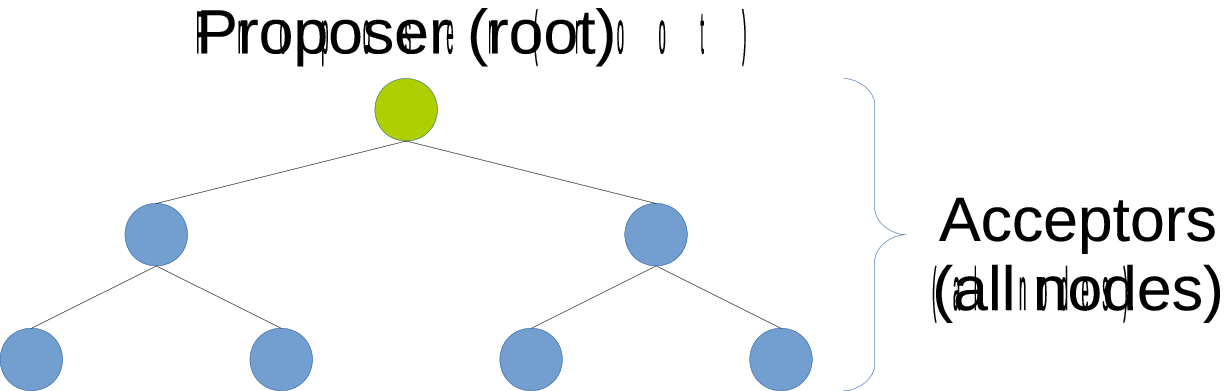}
\caption{...}
\label{fig:tree}
\end{figure}
}


\label{sec:tree-sign}



\com{Each \app protocol round consists of four phases,
As mentioned before, each node $i$ has a public key $X_i = G^{x_i}$,
where $x_i$ is $i$'s secret key
and $G$ is a generator of a suitable integer or elliptic curve group.
For each node $i$, we define the {\em composite} public key $\hat{X}_i$
as the combination of the public keys of node $i$ and all its descendants
using the group operation:
$\hat{X}_i = \prod_{j \in D_i} X_j = G^{\sum_{j \in D_i} x_j}$,
where $D_i$ is the set of transitive descendants of $i$ including $i$ itself.
Taking node $0$ to be the root,
each log entry produced by \app will be verifiable
using the network-wide composite public key $\hat{X}_0$,
hence producing these log entry signatures will require the participation of
every node $i$ with its ``share'' $x_i$ of the composite secret key.}

A single round of the \app protocol consists of four phases,
illustrated in Figure~\ref{fig:phases},
representing two communication ``round-trips''
through the leader-defined spanning tree:

\begin{compactenum}
\item
{\bf Announcement:}
The leader multicasts an announcement of the start of this round
down through the spanning tree,
optionally including the statement $S$ to be signed.

\item
{\bf Commitment:}
Each node $i$ picks a random secret $v_i$
and computes its individual commit $V_i = G^{v_i}$.
In a bottom-up process,
each node $i$ waits for an aggregate commit $\hat{V}_j$
from each immediate child $j$, if any.
Node $i$ then computes its own aggregate commit
$\hat{V}_i = V_i \prod_{j \in C_i} \hat{V}_j$,
where $C_i$ is the set of $i$'s immediate children.
Finally, $i$ passes $\hat{V}_i$ up to its parent,
unless $i$ is the leader (node 0).

\com{	simplify, one-step-at-a-time...
each node $i$ computes and sends its parent a message $(T_i,V_i,\hat{V}_i)$,
where $T_i$ is the root of an intermediate Merkle tree,
$V_i$ is a partial commitment for a collective signature, and
$\hat{V}_i$ is an aggregation of the partial commitments of
node $i$ and all its descendants.
Each node $i$ first waits to receive a commitment message
from each of its immediate children, if it has any.
Node $i$ then picks a random secret $v_i$,
computes $V_i = G^{v_i}$, and computes
$\hat{V}_i = V_i \prod_{j \in C_i} \hat{V}_j$,
where $C_i$ is the set of $i$'s immediate children.
Node $i$ now combines $V_i$, $\hat{V}_i$,
its log contribution for this round,
and the Merkle tree roots $T_j$
of each of its children $j \in C_i$,
to form $i$'s local log entry $L_i$ for this round.
A cryptographic hash of $L_i$ yields $T_i$,
the root of a new aggregate Merkle tree
representing the log contributions and commitments for this round
of $i$ and all its descendants.
Node $i$ finally sends $(T_i,V_i,\hat{V}_i)$ to its parent,
unless $i$ is the root node.
}

\item
{\bf Challenge:}
The leader computes a collective challenge
$c = \hash{\hat{V}_0 \parallel S}$,
then multicasts $c$ down through the tree,
along with the statement $S$ to be signed
if it was not already announced in phase 1.

\com{
The root node now uses its top-level Merkle tree hash $T_0$
along with the collective commitment $V_0$ and 
the hash of the current configuration policy $p$
as a challenge $c = H (T_0 \parallel V_0 \parallel p)$
with which all nodes will produce
their collective \sigt signature.
In a top-down communication process,
each node $i$ waits for a message from its parent,
then sends to each its immediate children $j \in C_i$:
(a) the top-level Merkle tree hash $T_0$ for this round, and
(b) the set of intermediate Merkle tree records on the path
between $T_0$ and $T_j$.
These tree records enable each child $j$ to verify efficiently
that its log contribution and commitments were correctly included
in the top-level aggregate Merkle tree for this round.
These records will also, later,
enable each node to prove to any third party
that its log contributions were included in this round.
}

\item
{\bf Response:}
In a final bottom-up phase,
each node $i$ waits to receive
a partial aggregate response $\hat{r}_j$
from each of its immediate children $j \in C_i$.
Node $i$ now computes its individual response
$r_i = v_i - c x_i$,
and its partial aggregate response
$\hat{r}_i = r_i + \sum_{j \in C_j} \hat{r}_j$.
Node $i$ finally passes $\hat{r}_i$ up to its parent,
unless $i$ is the root.

\com{
In a final bottom-up phase,
each node $i$ first waits for a message
from each of its immediate children $j \in C_i$, if any,
containing a partial aggregate response $\hat{r}_j$.
Node $i$ then computes its portion of the response
for the collective \sigt signature as
$r_i = v_i - c x_i$,
then computes
$\hat{r}_i = r_i + \sum_{j \in C_j} \hat{r}_j$.
Node $i$ finally passes $\hat{r}_i$ on up to $i$'s parent
unless $i$ is the root.
}
\end{compactenum}

The round announcement in phase 1 may, but need not necessarily,
include the statement $S$ to be signed.
Including $S$ in the announcement enables witnesses
to start validating the statement earlier
and in parallel with communication over the tree.
This approach is likely preferable when
witnesses may need significant time to validate the statement $S$,
such as when reproducing software builds
as an extreme example~\cite{bobbio14reproducible}.
On the other hand,
proposing $S$ later in phase 3
enables the leader to ``late-bind'' its statement,
perhaps incorporating information gathered from witnesses in phase 2,
as our timestamp service does (Section~\ref{sec:impl:time}).
Further, keeping phases 1--2 independent of the statement to be signed
in principle allows these phases to be performed offline ahead of time,
though we have not implemented or evaluated this offline variation.

During phase 4,
each node $i$'s partial aggregate response $\hat{r}_i$,
together with the collective challenge $c$,
forms a valid Schnorr multisignature on statement $S$,
verifiable against $i$'s partial aggregate commit $\hat{V}_i$ and
corresponding partial aggregate public key $\hat{X}_i$.
Anyone may compute $\hat{X}_i$
simply by multiplying the well-known public keys of $i$
and all of its descendants in the spanning tree.
Thus, each node can immediately check
its descendants' responses for correctness,
and immediately expose any participant producing an incorrect response.
\com{
Additionally, a parent node $i$ can use $V_j$ received along with 
with $\hat{V}_j$ to produce an exception in case of $j$'s failure 
in a way that does not necessitate removing $j$'s children,  
as detailed in Section~\ref{sec:excep}.
}
While nothing prevents a malicious node $i$
from computing $\hat{V}_i$ dishonestly in phase 2,
$i$ then will be unable to produce a correct response in phase 4
unless it knows the discrete logarithm $v_i$
such that $\hat{V}_i = G^{v_i}\prod_{j \in C_i} \hat{V}_j$.

The final collective signature is $(c,\hat{r}_0)$,
which any third-party may then verify as a standard Schnorr signature
by recomputing $\hat{V}'_0 = G^{\hat{r}_0}\hat{X}_0^c$
and checking that $c \stackrel{?}{=} \hash{\hat{V}'_0 \parallel S}$.
The scheme's correctness stems from the fact that
$\hat{V}_0 = G^{\sum_i v_i}$,
$\hat{r}_0 = \sum_i v_i - c \sum_i x_i$,
and
$\hat{X}_0 = G^{\sum_i x_i}$.
The scheme's unforgeability stems from the fact that
the hash function makes $c$ unpredictable with respect to $\hat{V}_0$,
and the collective cannot produce the corresponding response $\hat{r}_0$
without the (collective) knowledge of the secret key $x_i$
of {\em every node} $i$ whose public key
is aggregated into $\hat{X}_0$.
These properties are direct implications of the structure of Schnorr signatures,
which have been formally analyzed
in prior work~\cite{micali01accountable,bellare06multi},
though we are not aware of prior systems that
used these properties in practice to build scalable signing trees.

\com{
During phase 4,
each node $i$ may interpret $(c,\hat{r}_i)$ a \sigt signature
on the round's top-level Merkle tree $T_0$,
verifiable against the partial composite public key $\hat{K}_i$
comprising $i$'s public key and those of its descendants.
Once phase 4 completes,
the root node obtains a complete \sigt signature $(c,\hat{r}_0)$
verifiable against the top-level composite public key $\hat{K}_0$,
which aggregates the public keys of all nodes in the network.
The root node then simply broadcasts this top-level signature
to all nodes in order to ``complete'' the log entry for this round.
In practice this signature broadcast may be combined with
the announcement of the next round (\ie, the next round's phase 1).

\paragraph{Signature Verification} 
In order to verify the collective signature $(c,\hat{r})$ on a log entry $T_0$
with respect to a configuration policy $p$, a user performs the following steps.
First, the user retrieves public keys of all signatories listed in $p$ and calculates the 
composite public key $\hat{K}_0 = \prod_{j \in D_i} K_j$.
Then, the user recalculates the challenge used in the signature by first calculating
$V' = G^{\hat{r}} K_0^c$  and then $c' = H(T_0 \parallel V' \parallel p)$. The signature is accepted only if
$c'  = c$.  
}

\subsection{Accounting for Unavailable Witnesses}
\label{sec:live}
\label{sec:excep}

Authorities are unlikely to deploy witness cosigning
if their own availability may be degraded,
or even deliberately DoS-attacked,
by the unreliability of one or more witnesses.
We expect authorities to accept only witnesses
operated by reputable and competent organizations
who can normally be expected to keep their witness servers highly available,
so we expect the operational common case to be
for all witnesses to be present,
and only rarely for one or a few to be missing.

Unlike secret-sharing protocols~\cite{feldman87practical,stadler96publicly},
\cosi allows the leader to proceed
with {\em any} number of witnesses missing,
and merely documents these missing witnesses as {\em exceptions}
as part of the resulting collective signature.
Signature verifiers learn both how many
and {\em which} witnesses were missing
when an authoritative statement was signed,
and can independently determine their acceptance thresholds
via arbitrary predicates (Section~\ref{sec:sig-verify}).
The leader might set its own threshold as well:
\eg, if many or most witnesses are unreachable,
this may indicate the leader itself
is disconnected from much of the Internet,
making it useless and perhaps counterproductive
to sign further statements until connectivity is restored.

\com{
In the first approach,
we assume that failures of \app signers
are rare and brief,
such that at most a few nodes, \eg, $O(\log N)$,
are expected to be offline at any given time.
This stability might be satisfied by administrative fiat, for example:
\eg, the existing members of a \app community
might admit new members only after the candidate has demonstrated
adequate provisioning and high availability.
Individual participants might ensure this high availability, for example,
through conventional state machine replication such as 
Paxos~\cite{lamport98parttime,ongaro14search} or BFT~\cite{castro99practical}
across a small number of centrally-managed physical machines.
Such replication would operate below and be invisible to the \app protocol,
so we do not delve into the details here.

If server failures are rare but do occasionally happen,
we can account for a few failures in any given round by relaxing the demand
that each collective log entry be signed with the combination
of {\em all} nodes' public keys.
}

We start with a simple approach to handling witness failures,
then subsequently explore variations and optimizations.
In any of the phases of the tree-based signing protocol described above,
if any participant $i$ finds that one of its children $j$ is unreachable,
$i$ simply returns an error indicating the missing witness,
which propagates back up the tree to the leader.
The leader then reconfigures the tree to omit the missing witness,
announces the new tree,
and restarts the signing round from phase 1 over the new tree.
The leader includes in the resulting signature
not only the challenge and aggregate response $(c, \hat{r}_0)$
but also an indication of which witnesses were missing.
Verifiers then check the resulting signature
against a modified aggregate public key $\hat{X}$
computed by multiplying only the public keys of witnesses
that were actually present in the signing tree
(and hence contributed to the aggregate commit in phase 2
and the aggregate response in phase 4).

An intermediate witness in the leader's spanning tree
could maliciously pretend that one of its children is unavailable,
or a pair of witnesses might simply be unable to communicate
due to Internet routing failures.
To address this risk,
when a witness is reported ``missing''
the leader can first try contacting it directly
and/or request that other witnesses attempt to contact it.
If successful, the leader can then
reconnect the orphaned witness at a different location in the new tree.

\xxx{	from JB: lay out security argument for this more clearly. }

\subsection{Representing Exceptions in Signatures}
\label{sec:live:rep}

To minimize the size of collective signatures,
\cosi permits exceptions to be represented in three different ways:
as a list of witnesses absent, a list of witnesses present,
or a bitmap with one bit per witness.
After completing a signing round,
the leader simply chooses whichever representation
yields the smallest signature.
Listing witnesses absent yields the most compact signature
(less than 100 bytes using the Ed25519 curve~\cite{bernstein12highspeed})
in the hopefully common case when nearly all witnesses cosign.
Listing witnesses present is optimal at the opposite extreme,
while the bitmap approach is most efficient
in the region between those extremes.
Worst-case signature size is therefore about $2K+W/8$ bytes,
where $K$ is the size of a private key (\eg, 32 bytes for Ed25519)
and $W$ is the total number of witnesses,
plus a few encoding metadata bytes.

A more sophisticated alternative we explored
is to represent the witness roll call
as a Bloom filter~\cite{bloom70space},
which can sometimes increase compactness
at the risk of introducing false positives.
The leader might tolerate this false positive risk
by removing the contributions of falsely-marked witnesses
from the aggregate signature,
or salt the Bloom filter's hash functions
and ``mine'' to find a Bloom filter yielding no false positives.
We simulated several such approaches,
but did not find the results to be worth the additional complexity.


%


\subsection{Proactive, Retroactive, and Adaptive Validation}
\label{sec:sign:valid}

As discussed earlier in Section~\ref{sec:sign:witnesses},
the primary responsibility of witnesses is merely to
ensure proactively that signed authoritative statements are public --
but witnesses can and ideally should also check
the syntactic and semantic validity of statements when possible.
Some such validation checks may be feasible in principle
but require additional network communication
or take unpredictable amounts of time.

As one example, a witness to the signing of
a stapled OCSP certificate status~\cite{rfc6961}
or a CONIKS public key directory~\cite{melara15coniks}
might wish to verify that the certificates in these statements
are indeed fresh,
and are not listed in publicly available 
Certificate Revocation Lists (CRLs)~\cite{liu15end}.
If the witness were to initiate the 
fetching and downloading of CRLs
on the ``critical path'' of witnessing and cosigning, however,
then the witness might seriously delay the signing process,
or cause the leader to timeout
and consider the witness to have failed (Section~\ref{sec:live}).
To avoid such delays,
instead of fetching CRLs on the critical cosigning path,
certificate witnesses might periodically download
and maintain cached copies of relevant CRLs,
and merely check proposed OCSP staples or key directories
against their most recently cached CRLs.

Validation may sometimes be quick
but other times may require significant amounts of time
and/or computational resources.
A witness to a software update authority for an open source package,
for example (Section~\ref{sec:bg:update}),
might wish to verify the platform-specific binaries to be signed
against a reproducible build~\cite{porup15how}
of a corresponding source release in a public repository.
In this case, the witness may have to perform
an entire build of a large software tree before signing.
This delay may be acceptable in the special case of software updates,
which tend to be released on slow, latency-tolerant timescales anyway,
but such delays may not be acceptable in many other witnessing scenarios.

As one way of handling long or unpredictable validation delays,
the leader might specify a maximum validation time.
Each witness launches its validation process in parallel
but monitors it dynamically 
to see whether it actually completes in the required time.
If not, the witness might just ``cosign anyway,''
giving the leader the benefit of the doubt,
but continue the checking process and raise an alarm
in the hopefully rare event that validation eventually fails.
This approach of course weakens \cosi's transparency model
to be only ``proactive sometimes'' and ``retroactive sometimes.''
To create a public record of this distinction,
leaders might obtain two collective signatures
in parallel from all witnesses:
the first merely attesting that the witness has {\em seen} the statement,
and the second attesting that the witness has {\em validated} it.
Witnesses then provide the former cosignature but withhold the latter
if they cannot complete their validation in the time available.

\subsection{Limitations, Tradeoffs, and Future Work}
\label{sec:limits}

The most important limitation of witness cosigning
is that it requires active communication --
and perhaps {\em global} communication
if the witness group is highly distributed --
on the signing path.
This is a basic cost of \cosi's proactive approach to transparency:
by eliminating the need for the clients receiving an authoritative statement
to communicate at verification time
as gossip-based transparency approaches do~\cite{rfc6962, laurie14CT},
we incur the cost of communicating {\em before}
the authority's statement is made available to clients.

Because of the communication cost incurred at signing time,
\cosi is more suitable for authoritative signing activities
that can be done only periodically or in periodic batches,
and less suited to signing activities that must be done
individually in high volumes or at low latencies.
Fortunately, many authoritative signing activities
are already or can easily be performed periodically in batches.
For example, Section~\ref{sec:impl:time} presents a timestamp authority
that handles heavy client request loads by signing batches of timestamps,
and logging services such as CT's~\cite{laurie14CT},
as well as blockchains
used in cryptocurrencies~\cite{nakamoto08bitcoin,kokoris16enhancing},
routinely aggregate many client-requested transactions into large
latency-insensitive batches.

A second limitation of \cosi's approach
is that an authority's witness group cannot be completely ``open''
for anyone to join,
without making the system vulnerable to Sybil attacks~\cite{douceur02sybil}
in which an adversary creates and joins
a threshold number of colluding, fake witnesses.
One advantage of retroactive gossip-based checking~\cite{nordberg15gossiping}
is that ``anyone can gossip'' --
\ie, no entry barrier at all need be imposed
on the group of gossiping participants.
Thus, \cosi may best be viewed as complementary to
rather than a replacement for retroactive gossip-based consistency checking:
\cosi provides proactive security grounded in
a potentially large and diverse but at least somewhat selective witness group,
whereas gossip provides only retroactive protection
dependent on active communication but among a completely open
group of participants.

\section{Design Variations and Tradeoffs}
\label{sec:var}

While we expect the basic \cosi design described above
to be usable and suitable in many contexts,
as the evaluation in Section~\ref{sec:evals} suggests,
many improvements and design variations are possible
embodying different strengths and weaknesses.
We now briefly sketch some of this design space,
focusing on signature verification predicates,
reducing the size of the certificates
needed to verify collective signatures,
and tolerating
unreliability in the network and/or witnesses.

\subsection{Collective Signature Verification Predicates}
\label{sec:sig-verify}

\com{PJ: Wonder if that shouldn't be moved to the ``Design Variations and
Tradeoffs'' section?}	

Because \cosi signatures explicitly document
which witnesses did and did not participate in signing,
signature verification need not be based on a simple threshold,
but can in principle be an arbitrary predicate on subsets of witnesses.
For example, if the authority has reason
to trust some witnesses more than others,
then signature verification may be weighted
so that some witnesses count more than others toward the threshold.
To save signature space,
the authority can treat itself as a special ``witness,''
aggregating its own signature with all the others,
but imposing the rule that its own participation is mandatory
for the collective signature to be accepted.

Witnesses might be divided into multiple groups
with hierarchical expressions defining their relationships.
For example, a global body of witnesses might be divided
into geopolitical regions (\eg, Five Eyes, Europe, etc.),
each with different witness group sizes and thresholds,
such that a threshold number of regions
must in turn meet their respective internal thresholds.
Such a structure could protect the authority and its users
from compromise or denial-of-service
even if some regions contain many more witnesses than others
and {\em all} witnesses in any sub-threshold set of regions collude.

Finally, collective signature verification
might use different predicates depending on verification context.
Consider a device manufacturer desiring protection
from possible government coercion to produce secretly backdoored
operating system updates~\cite{ford16apple,doctorow16using}.
The manufacturer may be averse to the risk, however slight,
that a sufficient number of its witnesses
might become unavailable or collude
to prevent the manufacturer from signing legitimate updates.
The manufacturer could design its devices to mitigate this risk
by demanding a high cosigning threshold (\eg, half) when verifying updates
downloaded automatically or installed while the device is locked,
but allowing updates with few or no cosignatures
if the user manually initiates the update with the device unlocked.

This way, in the hopefully unlikely event the manufacturer becomes
unable to meet the normal cosigning threshold
due to massive witness failure or misbehavior,
the manufacturer can instruct users to install the next update manually,
and revise its witness group as part of that update.
More importantly,
the knowledge that the manufacturer has this fallback available
should deter any deliberate misbehavior by witnesses,
\eg, extortion attempts,
which would present only a minor inconvenience to the manufacturer's users
while likely yielding a public scandal
and lawsuits against the misbehaving witnesses.

\subsection{Reducing Authority Certificate Size with Key Trees}
\label{sec:certsize}

The basic \cosi design keeps collective signatures compact,
but requires that the authority's well-known certificate --
which verifiers need to check collective signatures --
include not just the authority's own public key but also
a complete list of the authority's witnesses and their public keys.
This large certificate size is acceptable
if it is distributed as part of a much larger package anyway,
\eg, embedded in a web browser's built-in root certificate store.
Large certificates
might be a problem in other contexts, however:
\eg, if they must be embedded in intermediate certificates,
DNSSEC~\cite{rfc4033} resource records, 
or other objects that are frequently transmitted.

In an alternate design yielding different tradeoffs,
the authority's certificate includes only
the authority's own public key,
the product of {\em all} witnesses' public keys
$\hat{X} = \prod_i X_i$,
and a hash representing the root of a {\em key tree}:
a Merkle tree~\cite{merkle88digital}
whose leaf nodes contain the individual witnesses' public keys.
The key tree hash in the authority's certificate
represents a universally-verifiable commitment
to all witnesses' public keys,
without the certificate actually containing them all.

During subsequent signing rounds,
the \cosi leader includes in each signature
a list of the public keys of all missing or present witnesses,
whichever is shorter,
along with Merkle inclusion proofs for each
proving their presence in the authority's key tree.
To check a signature containing a list of present witnesses,
the verifier simply multiplies the listed public keys
(after verifying their inclusion proofs).
To check a signature containing a list of missing witnesses,
the verifier multiplies the aggregate $\hat{X}$ of all witnesses' public keys
with the inverses of the missing witnesses' public keys:
$\hat{X}' = \hat{X} \prod_{j \in L} X_j^{-1}$.

In the hopefully common case in which all witnesses are present during signing,
the signature is at minimum size, containing only $(c,\hat{r}_0)$
and an empty list of missing witnesses.
As more witnesses go missing, however,
the size of the signature including witness public keys and inclusion proofs
may grow to $O(N)$ size,
or potentially even $O(N \log N)$
if each missing witness's inclusion proof is stored separately
without sharing the storage of internal key tree nodes.

\subsection{Gracefully Tolerating Network Unreliability}
\label{sec:netchurn}

While we expect authorities adopting \cosi
to choose reliable witness servers run by reputable organizations,
neither the authority nor its witnesses can control
the Internet connections between them.
\cosi allows the authority
to rebuild its communication trees at any time
to route around link failures,
but if network churn is too frequent or severe,
a tree might become unusable before it can be used even once.

One attractive solution to this problem is to adopt
the {\em binomial swap forest} technique
of San Ferm\'in~\cite{cappos08sanfermin},
which is readily applicable to \cosi.
We first assign all witnesses $b$-bit binary labels.
We then implement each of \cosi's aggregation rounds --
\ie, its Commit and Response phases --
with a single run of San Ferm\'in's dynamic aggregation protocol.
To aggregate commits or responses,
each node communicates with $b$ other nodes in succession,
building up its own aggregate while simultaneously
helping other nodes build theirs,
such that {\em every} participant ends up obtaining a complete aggregate.

At each swap step $i$ from $0$ to $b-1$,
each witness $j$ communicates with another witness $k$ whose label
differs at bit $i$ but is identical in all more-significant bits.
At step $0$,
each even-numbered node swaps with its immediate odd-numbered neighbor.
During subsequent steps, however, each witness has a choice
of witnesses to swap with:
\eg, in step $1$ a node labeled $xx00$ may swap with either $xx10$ or $xx11$.
In these swaps each witness combines the other witness's aggregate value
from prior steps into its own aggregate,
enabling both communication partners to double the ``coverage''
of their respective aggregates in each step,
until every witness has a complete aggregate.
The authority may then pick up this complete aggregate --
\ie, the collective commit or response in the case of \cosi{} --
from any witness server.

Because each witness can dynamically choose its
communication partners in steps $i>0$,
witnesses can adapt immediately to intermittent link failures
without restarting the overall aggregation process,
provided the witnesses themselves do not fail.
Tolerating high churn in the witnesses as well as the network
requires other techniques explored below.

\subsection{Avoiding Signing Restarts on Witness Unreachability}
\label{sec:restarts}

A second-order availability risk in the basic \cosi design
is that multiple witnesses might become unavailable
during a single signing round --
perhaps even intentionally as part of a DoS attack by malicious witnesses --
thereby forcing the leader to restart the signing round
multiple times in succession without making progress.
To address this risk we may prefer
if the leader could always complete each signing round,
and never have to restart,
regardless of the witnesses' behavior.

If during \cosi's Commit phase
some witness $i$ finds
one of its immediate children $j \in C_i$ unresponsive,
$i$ can adjust its aggregate commit $\hat{V}_i$
to include only its own individual commit $V_i$
and the aggregate commits of its children who {\em are} reachable,
and pass the adjusted $\hat{V}_i$ to $i$'s parent
along with a list of unreachable witness(es).
The signing round can thus immediately
take the missing witnesses into account
and continue without restarting.
If a missing witness $j$ is an interior node in the spanning tree,
then its parent $i$ (or the leader)
can attempt to ``bridge the gap'' by contacting $j$'s children directly
to collect their portions of the aggregate commitment
(and their corresponding portions of the aggregate response later in phase 4).
Thus, the loss of an interior node in the spanning tree
need not entail the loss of its descendants' cosignatures.

A more subtle challenge occurs
when some witness $j$ participates in the Commit phase
but goes offline before the subsequent Response phase.
In this case, the missing witness's individual Schnorr commit $V_j$
has been included in the aggregate commit $\hat{V}_0$
and used to form the collective challenge $c = \hash{\hat{V}_0 \parallel S}$
with which all witnesses must compute their collective responses.
Thus, it is now too late to change $c$,
but without witness $j$ the remaining witnesses
will be unable to produce an aggregate response $\hat{r}_0$
matching the aggregate commit $\hat{V}_0$ that included $j$'s commit.
Further, breaking the dependency of $c$ on $\hat{V}_0$ --
allowing the latter to change in the Response phase
without recomputing $c$ --
would make the collective signature trivially forgeable.

\com{ from Cappos:
I got lost in the signature details in the paper, so I did not validate that part of the scheme.  Later, when you talk about "all the possible aggregate commits" in III.I, I am lost by what you mean.
}

We can resolve this dilemma
by making the collective challenge $c$ depend
not on just a single aggregate commit $\hat{V}_0$ of individual commits $\hat{V}_i$
but on {\em all possible} aggregate commits $\hat{V}_W$
representing any subset of the witnesses $W$
that participated in the Commit phase.
During the Commit phase, these witnesses
no longer merely aggregate their individual Schnorr commits,
but also include them in a Merkle tree summarizing all individual commits.
Each interior witness $i$ obtains from each of its children $j \in C_i$
both $j$'s aggregate commit $\hat{V}_j$
and the hash $H_j$ representing a partial Merkle tree
summarizing all the individual commits of $j$'s descendants.
Then $i$ computes its aggregate as before,
$\hat{V}_i = V_i \prod_{j \in C_i} \hat{V}_j$,
but also produces a larger Merkle commit tree whose hash $H_i$
contains both $V_i$ as a direct leaf
and all of $i$'s childrens' Merkle commit trees $H_{j \in C_i}$ as subtrees.
The leader in this way obtains a root hash $H_0$
summarizing all witnesses' individual commitments,
and computes the collective challenge to depend on
the root of this commit tree,
$c = \hash{\hat{V}_0 \parallel H_0 \parallel S}$.

Now, in the hopefully common case
that all witnesses present in the Commit phase
remain online through the Response phase,
the witnesses produce an aggregate response $\hat{r}_0$ as before,
which matches the complete aggregate commit $\hat{V}_0$
appearing directly in the challenge.
If witnesses disappear after the Commit phase, however,
the leader includes in its signature
the individual commits of the missing witnesses,
together with Merkle inclusion proofs demonstrating that
those individual commits were fixed
before the collective challenge $c$ was computed.
The verifier then multiplies the aggregate commit $\hat{V}_0$
with the inverses of the individual commits of the missing witnesses,
to produce an adjusted aggregate commit $\hat{V}'_0$
and corresponding aggregate response $\hat{r}'_0$.

\subsection{Extreme Witness Churn and Asynchronous Networks}
\label{sec:async}

Schnorr signatures are well-established
and compatible with current best practices for standard digital signatures,
but their $\Sigma$-protocol nature (commit, challenge, response)
has the drawback of requiring two communication round-trips
through a distributed structure --
whether a simple tree or a binomial swap forest --
to aggregate a collective signature.
This requirement could be limiting in highly unstable or asynchronous situations
where any distributed structure built in the first round-trip
might become unusable before the second.

BLS signatures~\cite{boneh01short} may offer
an appealing alternative cryptographic foundation for \cosi,
requiring pairing-based elliptic curves
but avoiding the need for two communication round-trips.
In short, a BLS public key is $G^x$ as usual,
but a BLS signature is simply $\hash{M}^x$,
where $\hash{M}$ is a hash function mapping the message $M$
to a pseudorandom point on the appropriate curve.
Signature verification uses the pairing operation to check that
the same private key $x$ was used in the public key and the signature.
BLS extends readily to multisignatures,
since an aggregate signature $\hash{M}^{x_1+\dots+x_n}$
is simply the product of individual signatures $\prod_{i=1}^n{\hash{M}^{x_i}}$
and is verifiable against an aggregate public key $G^{x_1+\dots+x_n}$
computed in the same fashion as $\prod_{i=1}^n{G^{x_i}}$.

Using BLS instead of Schnorr signatures,
an authority can produce a collective signature
in a single round-trip through a tree or binomial swap forest
(Section~\ref{sec:netchurn}),
eliminating the risk of a witness participating in the commit phase
but disappearing before the response phase (Section~\ref{sec:restarts}).
Further, BLS signatures may make \cosi usable in protocols designed for
asynchronous networks~\cite{cachin00random,cachin01secure,ramasamy06parsimonious}
by allowing participants to aggregate signatures incrementally
and make use of them as soon as an appropriate threshold is reached:
\eg, typically $f+1$ or $2f+1$ in asynchronous Byzantine consensus protocols
tolerating up to $f$ faulty participants.

One key challenge in fully asynchronous aggregation,
where participants must dynamically adapt to arbitrary delay patterns,
is that nodes must be able to combine potentially overlapping aggregates
without imposing regular structures as used in San Ferm\'in.
For example, nodes $A$ and $B$ may communicate to form aggregate $AB$,
nodes $B$ and $C$ then form aggregate $BC$,
and finally nodes $A$ and $C$ must combine aggregates $AB$ with $BC$.
Aggregating BLS signatures as usual here
will yield a collective signature $\hash{M}^{x_A+2x_B+x_C}$
in which $B$'s signature is effectively aggregated twice.
There is no readily apparent way to avoid such duplication,
apart from just keeping the individual signatures separate
and giving up the efficiency benefits of incremental aggregation.

Such duplication may be tracked and compensated for, however,
by maintaining with each aggregate a vector of coefficients
indicating the number of ``copies'' of each node's signature (possibly 0)
represented in a given aggregate.
Thus, the aggregate $AB^2C$ from the above example would be represented
by the curve point $\hash{M}^{x_A+2x_B+x_C}$
and the coefficient vector $v=[1,2,1]$.
The number of participants represented in a given aggregate
is simply the number of nonzero elements in the coefficient vector.
Signature verification uses the coefficient vector to compute
the corresponding aggregate public key against which to verify the signature,
as $\prod_{i=1}^n{(G^{x_i})^{v_i}}$.
This approach has the downside of requiring $O(N)$ communication cost
per aggregation step due to the need to transmit the vector,
and $O(N)$ computation cost to compute
the correct aggregate public key in signature verification.
Partly mitigating these costs, however,
the vector's elements are small (\eg, one or two bytes)
compared to full elliptic curve points representing individual signatures,
and group exponentiation (scalar multiplication of curve points)
with small non-secret values can be made relatively inexpensive computationally.

\section{Prototype Implementation}
\label{sec:impl}

We have built and evaluated a working prototype witness cosigning \coty,
implementing the basic \cosi protocol described in Section~\ref{sec:sign}.
The prototype also
demonstrates \cosi's integration into two different authority applications:
a timestamp service, and a backward-compatible
witness cosigning extension
to the Certificate Transparency log server.


The \cosi prototype is written in Go~\cite{golang};
its primary implementation consists of 7600 lines of server code
as measured by CLOC~\cite{cloc}.
The server also depends on a custom 21,000-line 
Go library of advanced crypto primitives
such as pluggable elliptic curves,
zero-knowledge proofs, and verifiable secret sharing;
our \cosi prototype relies heavily on this library
but does not use all its facilities.
Both the \cosi prototype and the crypto library
are open source and available on GitHub:

\begin{center}
\texttt{\url{https://github.com/dedis/cothority}}
\end{center}


The \coty prototype currently implements
tree-based collective signing as described above
in Section~\ref{sec:sign}
including the signing exception protocol for handling witness failures.

We evaluated the \coty implementation with Schnorr signatures
implemented on the Ed25519 curve~\cite{bernstein12highspeed},
although the implementation also works and has been tested with
other curves such as the NIST P-256 curve~\cite{ansi05ecdsa}.


\label{sec:app}

\com{
While this paper's primary technical focus is on the \cosi protocol
for collective signing,
we now explore ways it may be applied
to different types of authorities.
Building realistic, fully functional cothorities
in each of these application areas will require significant 
additional application-specific design and develop
beyond the basic \cosi ``engine'' developed in this paper,
but we attempt to sketch broad approaches and important considerations
for each of these application areas.

\xxx{	Time services like NTP?  How do I know the time I was given
	is actually fresh, or even near the true time,
	if someone is able to control my network access? }
}

\subsection{Witness Cosigned Time and Timestamp Service}
\label{sec:timestamp}
\label{sec:impl:time}

As one application of witness cosigning,
we built a digital timestamping service~\cite{haber91how,adams01internet,sirer13introducing},
which also doubles as a coarse-grained secure time service.
The primary timestamp server, serving as the \cosi leader,
initiates a new signing round periodically --
currently once every 10 seconds --
to timestamp a batch of documents or nonces submitted by clients.
While the timestamp server could initiate a fresh witness cosigning round
to service each client timestamping request,
this mode of operation would be unlikely to scale to serve
large timestamp request transaction rates,
due to the global communication \cosi imposes on each signing round
(see Section~\ref{sec:limits}).

\subsubsection{Timestamp Request Processing}

A client wishing to timestamp a document
opens a connection to the timestamp server
and submits a hash of the document to stamp.
Many clients can have outstanding timestamp requests at once,
and a single client can concurrently submit timestamp requests
for multiple documents at once;
the timestamp server enqueues these requests but does not answer them
until the next signing round has completed.
At the beginning of each signing round,
the timestamp server collects all of the hashes
submitted since the previous round
into a Merkle tree~\cite{merkle88digital},
and prepares a timestamp record to sign consisting of
the current time and the root of this round's timestamp tree.
The timestamp server does not actually log these timestamp records,
but the records are hash-chained together in case witnesses wish to do so.
The timestamp server uses \cosi
to distribute the new timestamp record to all available witnesses
and produce a collective signature on the timestamp record.

Finally, the timestamp server replies to the outstanding client requests,
giving each client a copy of the timestamp record and
a standalone inclusion proof relating the client's submitted hash
to the Merkle tree root contained in the time\-stamp record.
To verify that a document was indeed included,
the verifier of a document timestamp 
uses the document's hash,
the timestamp server's certificate
(including the public keys of all witnesses),
the timestamp record, and the Merkle inclusion proof,
to verify that the document was indeed timestamped in that round
and that a threshold number of witnesses validated the timestamp record.

The timestamp server never records or transmits the full Merkle tree itself,
and forgets the Merkle tree after the round concludes.
The server transmits only individual inclusion proofs
to satisfy client requests.
Thus, the timestamp server leaves to clients
the responsibility of remembering timestamp records
and cryptographic evidence that a particular document was timestamped.
The primary security property is bound into
the timestamp record's collective signature,
which attests that the witnesses verified that
the record was formed and signed at approximately
the time indicated in the timestamp record.

\subsubsection{Coarse-grained Time Checking}

Since the timeserver does not care whether a value submitted for timestamping
is  actually a hash of documents or merely a random number,
clients can submit a random nonce to timestamp a ``challenge''
and obtain a witness cosigned attestation of the current time.
Timestamping a challenge in this way ensures
that attackers cannot replay valid but old signed timestamp records
to trick clients into thinking the time is in the past:
the client can verify directly that the timestamp record is fresh,
and can trust the timestamp it contains
on the assumption that a threshold of the timestamp server's witnesses
are honest.

Such a coarse-grained time-check may be useful
as a sanity-check for the client's NTP sources~\cite{mills91internet,rfc5905},
enabling the client to protect itself against
both compromised NTP servers and other time-related
vulnerabilities~\cite{malhotra15attacking}.
Due to the coordination required for collective signing,
\cosi's coarse-grained time checking will not substitute
for fine-grained NTP-based clock synchronization.
\cosi's coarse-grained sanity checking is instead complementary to NTP,
increasing security and ensuring that clients cannot be tricked
into believing that the time is far removed from reality in either direction.

\subsubsection{Scalable Timestamping}

To illustrate how applications can further leverage \cosi's architecture
in application-specific ways,
we enhanced the timestamp server prototype
to enable the witnesses, in addition to the leader,
to serve timestamp requests submitted by clients.
Thus, all witnesses effectively become timestamp servers
and can distribute the task of handling heavy client timestamp loads.
In this use of \cosi,
the leader defers formation of the timestamp record to be signed
until the beginning of the Challenge phase (Section~\ref{sec:tree}).

During the Commitment phase,
each witness collects all timestamp requests clients
submitted since the last round into a local Merkle timestamp tree,
including the timestamp tree roots generated by child witnesses,
then passes the aggregated Merkle timestamp tree up to the witness's parent.
The leader thus forms a global timestamp tree
that transitively includes all witnesses' local timestamp trees.

During the Challenge phase,
the leader passes down to each witness
an inclusion proof relating the root timestamp record
to the root of the witness's local timestamp tree.
Once the \cosi signing round concludes, forming the collective signature,
each witness can compose its inclusion proof
with the inclusion proof for each client request
within its local timestamp tree,
to give each client a complete inclusion proof
relating that client's submitted hash with the signed timestamp record.

\com{
Free Timestamping Authority 
DigiStamp, the Online Notary, the Timestamp Authority  
Internet X.509 Public Key Infrastructure Time-Stamp Protocol (TSP) RFC3161 
ANSI ASC X9.95 Standard expands RFC3161

Two scenarios: 1) the leader proposes a new log entry which is endorsed by the entire \coty,
in which case some nodes cannot change the entries because other nodes sign against the initial entry, 
which is included in the challenge, 
or 2) each node proposes its own addition to the log entry, everything gets aggregated into one MT, and 
the nodes does no produce a signature on the top hash (everyone is affected) unless it's got a proof of inclusion.
Either way, this issues is more about completely unresponsive nodes (availability section). }

\subsection{Witness Cosigned Certificate Logging Service}
\label{sec:impl:log}

As a second application and test-case building on an existing service,
we incorporated \cosi as a backward-compatible extension to Google's existing
Certificate Transparency log server~\cite{rfc6962, laurie14CT}.
CT's log server periodically constructs
a Merkle tree of records for recently timestamped and logged certificates,
and creates a Signed Tree Head (STH) representing
the root of each epoch's tree of timestamp records.
With our extension,
the log server attaches a collective witness signature to each STH
alongside the log server's existing individual signature.
Since the collective signature is carried in an extension field,
legacy CT clients can simply ignore it,
while new CT clients that are aware the log server supports witness cosigning
can verify the witness signature extension.

CT normally relies on a gossip protocol~\cite{nordberg15gossiping}
to enable other {\em auditor} servers to check retroactively
that a log server is behaving correctly,
and not revising or forking its history for example.
Our extension effectively makes this auditing function proactive,
enabling the log server's witnesses to check the log server's behavior
{\em before} each record is signed
and withhold their cosignature on the STH if not.

The protection this extension offers CT clients in practice
depends of course on client behavior.
Current CT clients typically check only individually-signed log server
timestamp records attached to logged certificates,
and thus would not directly benefit from collective signatures on STHs.
A log server could in principle witness cosign each timestamp record,
but the communication cost could become prohibitive
for log servers that timestamp a high volume of certificates.

However, independent of our witness cosigning extension,
CT is currently being enhanced so that clients can
obtain from web servers
not only the appropriate timestamp record but
the STH representing the epoch in which it was logged,
and an inclusion proof demonstrating that the timestamp record
was included in the STH for the relevant epoch.
Thus, CT clients supporting both this STH inclusion proof extension
and our STH cosigning extension
can obtain proactive protection
from secret attacks by powerful adversaries who might
have compromised both a CA's key and a few log servers' keys,
and who might be able to block the client's active communication
with uncompromised log servers.

\com{
\subsection{Witnessed Vote-Tallying}

Finally, to test voting and group evolution mechanisms
we
extended signing nodes as to be able to vote on any matter by signing
individual votes
in addition to collectively signing all votes.
\xxx{this part of the sentencedangles}
The
combination of signatures allowed a node to verify all other votes
and the correctness of the tally.
While this verification process currently incurs $O(N)$ costs,
we expect it could be reduced with improved, verifiable tallying methods
in the future.

}


\section{Evaluation}
\label{sec:evals}

The primary questions we wish to evaluate
are whether \cosi's witness cothority architecture is practical
and scalable to large numbers, \eg, thousands of witnesses,
in realistic scenarios.
Important secondary questions are what the important costs are,
such as signing latencies and computation costs.

While this paper's primary focus is on the basic \cosi protocol
and not on particular applications or types of \coties,
we also evaluated the \cosi prototype 
in the context of the timestamping and log server applications discussed above.

\subsection{Experimental Setup}

We evaluated the prototype on DeterLab~\cite{deterlab},
using up to 32 physical machines configured in a star-shaped virtual topology.
To simulate larger numbers of \cosi participants
than available testbed machines,
we run up to 1,058 \cosi witness processes on each machine
to perform experiments
with up to 33,825 witnesses total, corresponding to a fully populated
tree of depth 3 and a branching factor of 32.
A corresponding set of \cosi client processes on each machine
generate load by issuing regular timestamp requests
to the server processes.

To mimic a conservatively slow, realistic wide-area environment in which 
the witness \coty's servers might be distributed around the world,
the virtual network topology imposes a round-trip latency of 200 milliseconds
between any two witnesses.
The witnesses aggregate timestamp statements
from their clients and request every second the batch of statements to
be signed collectively as part of a single aggregate Merkle tree per round.
These testbed-imposed delays are likely pessimistic;
global deployments could probably achieve lower latencies
using approximate shortest-path spanning trees.

\com{
Performance: latency, bandwidth, CPU cost for a single trusted leader, flat structure (depth 1 tree), and a regular Merkle tree. 
Reliability: producing signatures with and without failures (unavailable signatories).

Verification: in case of the beacon application, the verification is basically checking a single signature, slightly more
work if not all signatories were available;
in case of the timestamping service, in addition to verifying the signature we must first check that an entry was properly committed.
}

\subsection{Scalability to Large Witness Cothorities}

\begin{figure}[t]
\centerline{\includegraphics[width=0.45\textwidth]{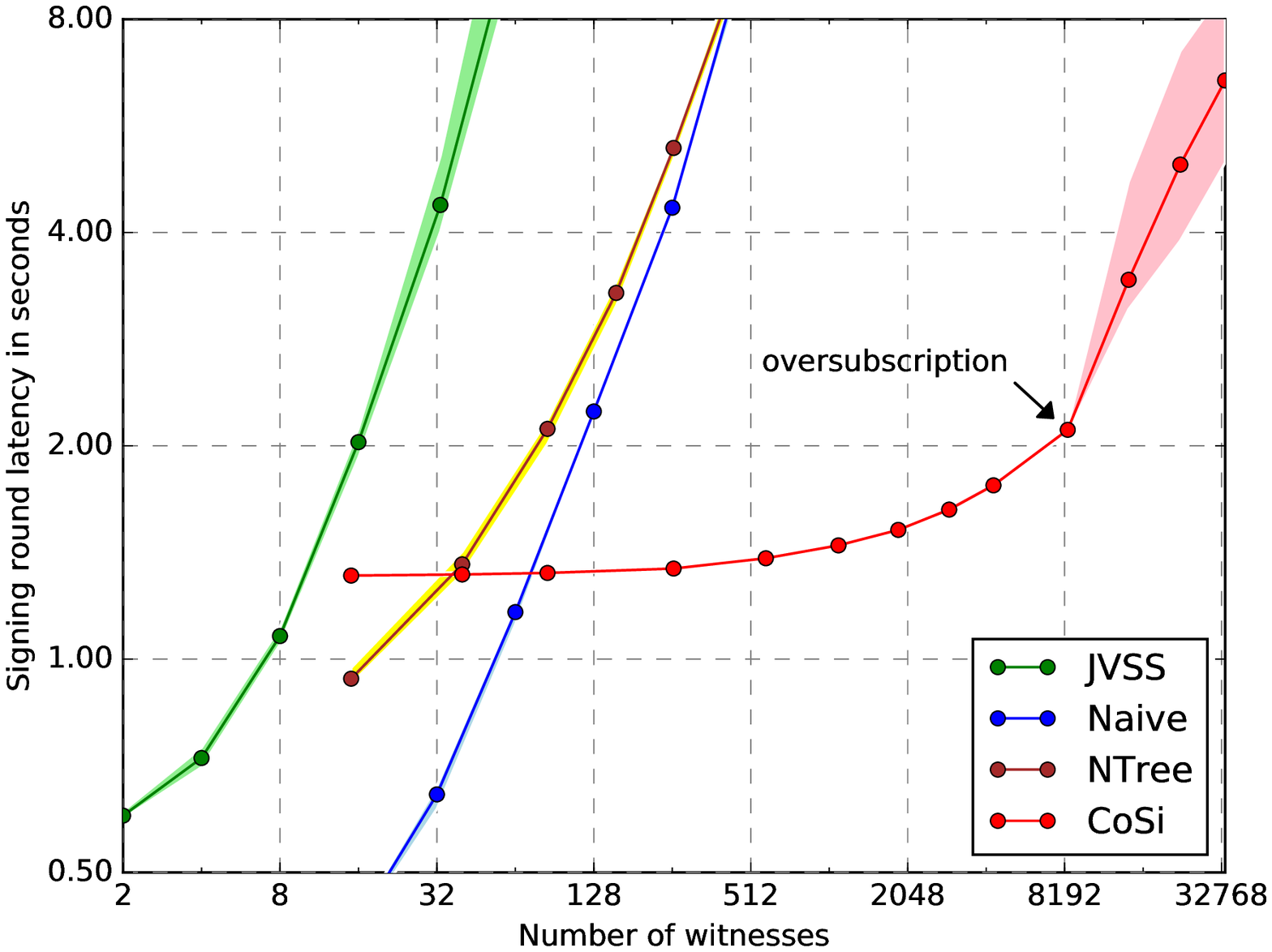}}
\caption{Collective signing latency versus number of participating witnesses.}
\label{fig:latency-vs-hosts}
\end{figure}

Our first experiment evaluates the scalability of the \cosi protocol
while performing simple collective signing rounds
across up to 33,825 witnesses.
We compare \cosi's performance against three different baselines.
The first baseline is ``Naive'' scheme in which the leader simply collects
$N$ standard individual signatures
via direct communication with $N$ witnesses.
Second, an ``NTree'' scheme still uses $N$ individual signatures, but
the $N$ witnesses are arranged in a communication tree and each node 
verifies all signatures produced within its subtree.
Finally, a ``JVSS'' scheme implements Schnorr signing using
joint verifiable secret sharing~\cite{feldman87practical,stadler96publicly}.

Figure~\ref{fig:latency-vs-hosts} shows
the results of this scalability experiment.
The lines represent averages measured over ten experimental runs,
while the shaded areas behind the lines represent the minimum and maximum
observed latencies over all ten runs.
\cosi's signing latency increases with the number of hosts as we would expect,
scaling gracefully with total number of witnesses
up to around 8,192 witnesses,
where the performance impacts of testbed oversubscription begin to dominate
as explored later in Section~\ref{sec:eval:oversub}.
Per-round collective signing latencies average slightly over 2 seconds
with 8,192 cosigning witnesses.
The maximum latency we observed in that situation was under 3 seconds over many runs.
Given that many authority protocols are
or can be made fairly latency-tolerant,
often operating periodically at timescales of minutes or hours,
these results suggest that witness cosigning
should be practical to enhance the security of many such authorities.

The Naive scheme
is naturally simpler and as a result faster for small witness groups,
but becomes impractical beyond around 256 witnesses
due to the costs of computing, transmitting, and verifying
$N$ individual signatures.

The even poorer performance of the NTree scheme can be traced back to the increasing
computational load each node must handle the further up it resides in
the communication tree.
As with the Naive scheme, NTree becomes impractical beyond
around 256 witnesses.

The JVSS approach proves to be the least scalable variant,
becoming impractical beyond about 32 witnesses.
This poor scalability results from the fact that JVSS
requires each of the $N$ witnesses to serve in a ``dealer'' role,
each producing an $N$-share secret polynomial
whose shares are encrypted and sent to the other $N$ nodes.
Every node must then combine the $N$ public polynomials
and the $N$ encrypted shares it receives
to form shares of a joint master polynomial.
In threshold Schnorr signing using JVSS,
this $O(N^2)$ dealing cost is incurred both
during initial key-pair setup and during {\em each} signing round,
because it is required to produce a fresh shared Schnorr commit 
$\hat{V}_0$ each round
whose private value is not known to any individual
or sub-threshold group of participants.
Using a pairing-based signature scheme such as BLS~\cite{boneh01short}
in place of Schnorr
could eliminate the need to deal a fresh commit per signing round
and thus reduce the per-round cost of JVSS signing,
but the $O(N^2)$ joint dealing cost would still be required
at key generation time.

\subsection{Computation Costs}

\begin{figure}[t]
\centerline{\includegraphics[width=0.45\textwidth]{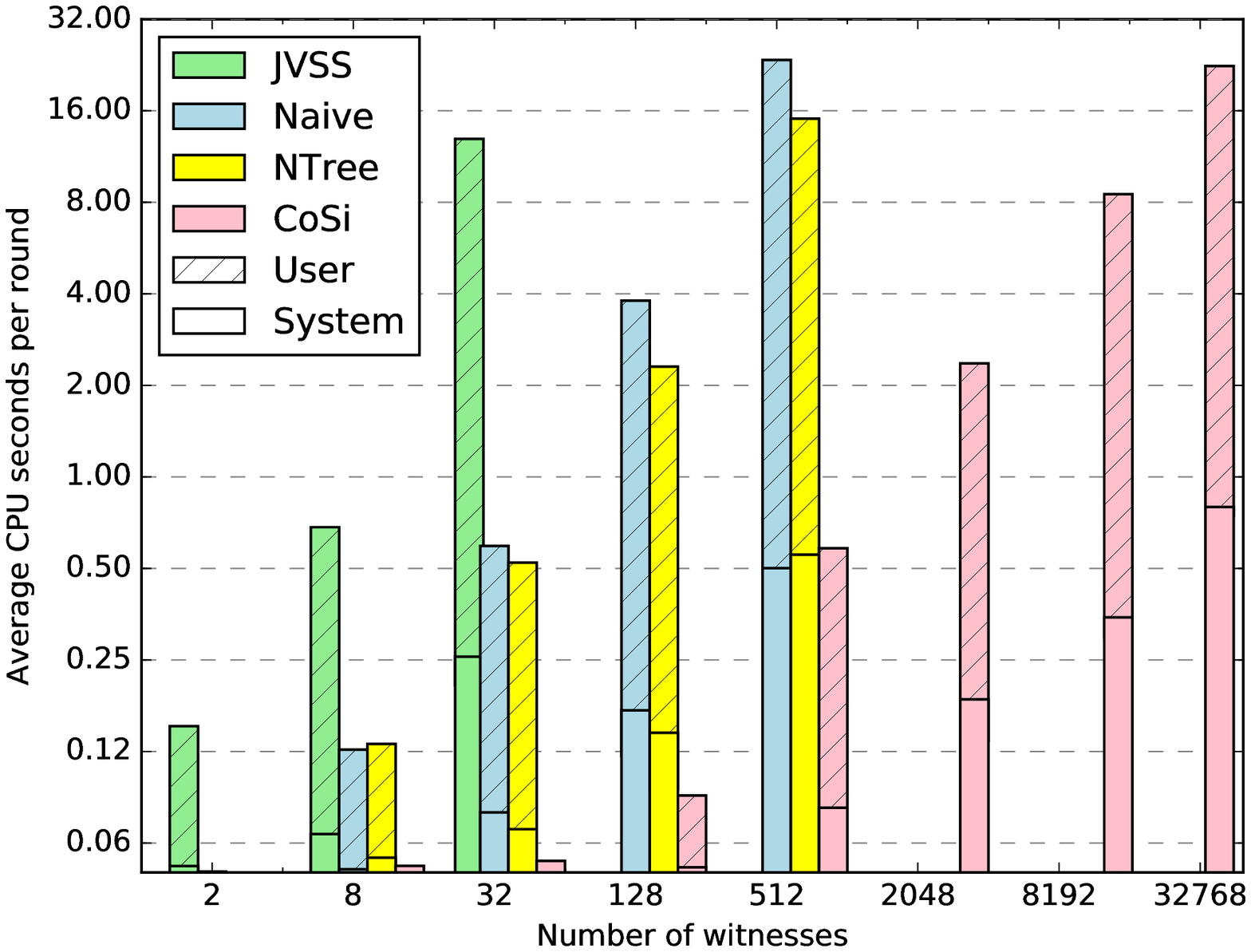}}
\caption{Per-node, per-round computation cost
	versus number of participating witnesses.}
\label{fig:compute-costs}
\end{figure}

The next experiment focuses on the protocol's per-node computation costs
for signing and signature verification.
The \cosi leader periodically initiates new collective signing rounds,
and we measure the total CPU time per round imposed on
the most heavily-loaded participant.
Since all \cosi participants check the (partial) signatures 
submitted by their children in the process of
producing the full aggregate signature,
this computation cost includes the cost of signature checking.

Figure~\ref{fig:compute-costs} shows how measured System and User time
on the most heavily-loaded signing node (typically the root)
varies depending on the number of cosigning witnesses.
The figure also shows the computation costs
of comparable Naive and NTree cosigning approaches using individual signatures,
as well as using joint verifiable secret sharing (JVSS).
As expected, the computational cost of the \cosi protocol
stays relatively flat regardless of scale,
whereas the computation costs of the competing schemes
begin to explode with groups beyond a few tens of witnesses.

The measured computation time is often greater than
the wall-clock signing latency because
computation is done in parallel and the graph represents the sum of
the CPU time spent by all threads running on a given witness server.

\subsection{Network Traffic}

\begin{figure}[t]
\centerline{\includegraphics[width=0.45\textwidth]{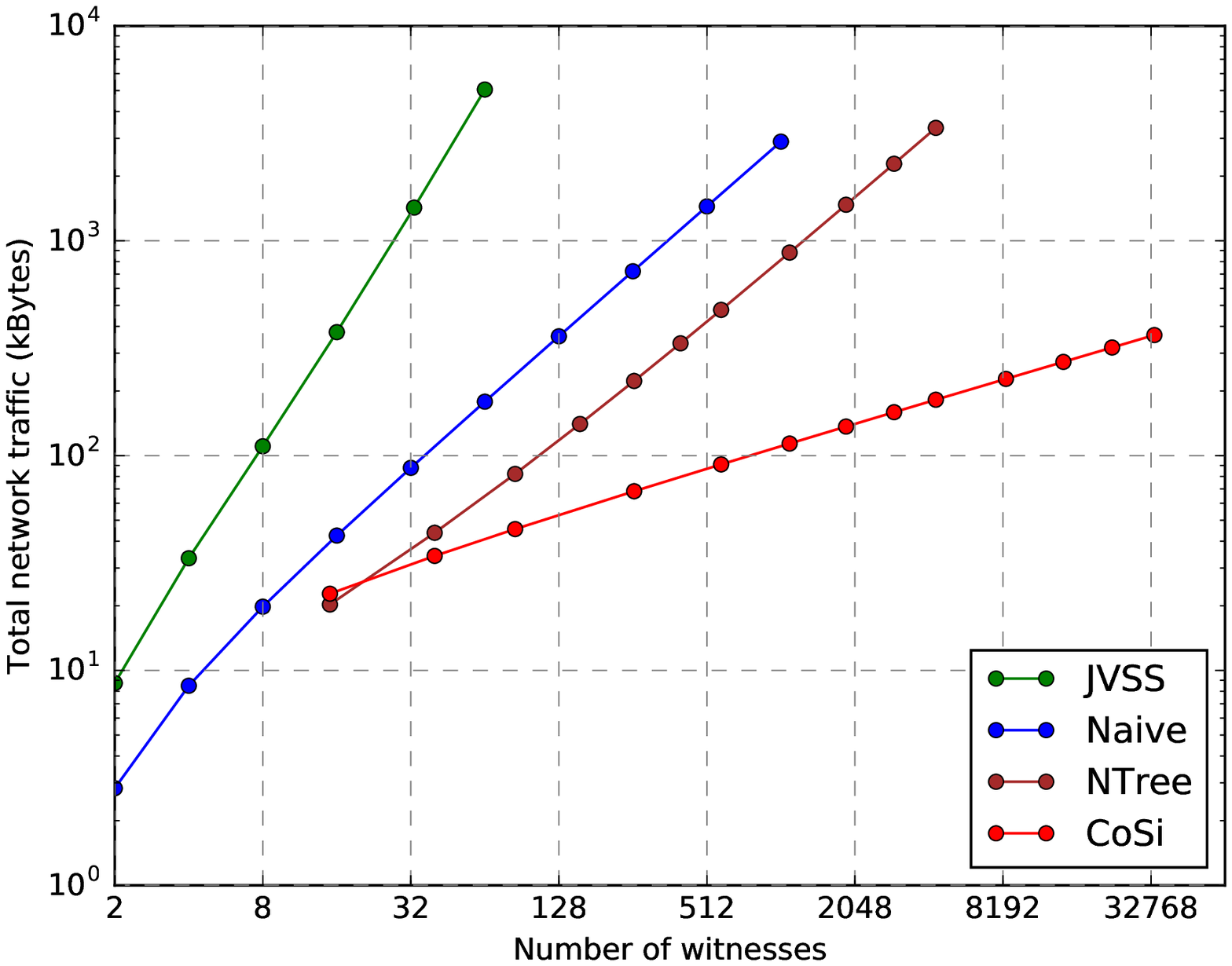}}
\caption{Network traffic (bandwidth consumption) at the root node
	versus number of participating witnesses.}
\label{fig:network}
\end{figure}

The next experiment measures the total network traffic produced by
CoSi in comparison with the Naive, NTree, and JVSS baselines.
Figure~\ref{fig:network} shows these results. Due to CoSi's
aggregation mechanism,
network traffic at the root node rises much more slowly
than in the the baseline schemes, which all lack the benefit of
aggregation, as the number of witnesses grows.
JVSS puts a particularly high burden on the network
due to its $O(N^2)$ communication complexity.

\subsection{Effects of Spanning Tree Configuration}

\begin{figure}[t]
\centerline{\includegraphics[width=0.45\textwidth]{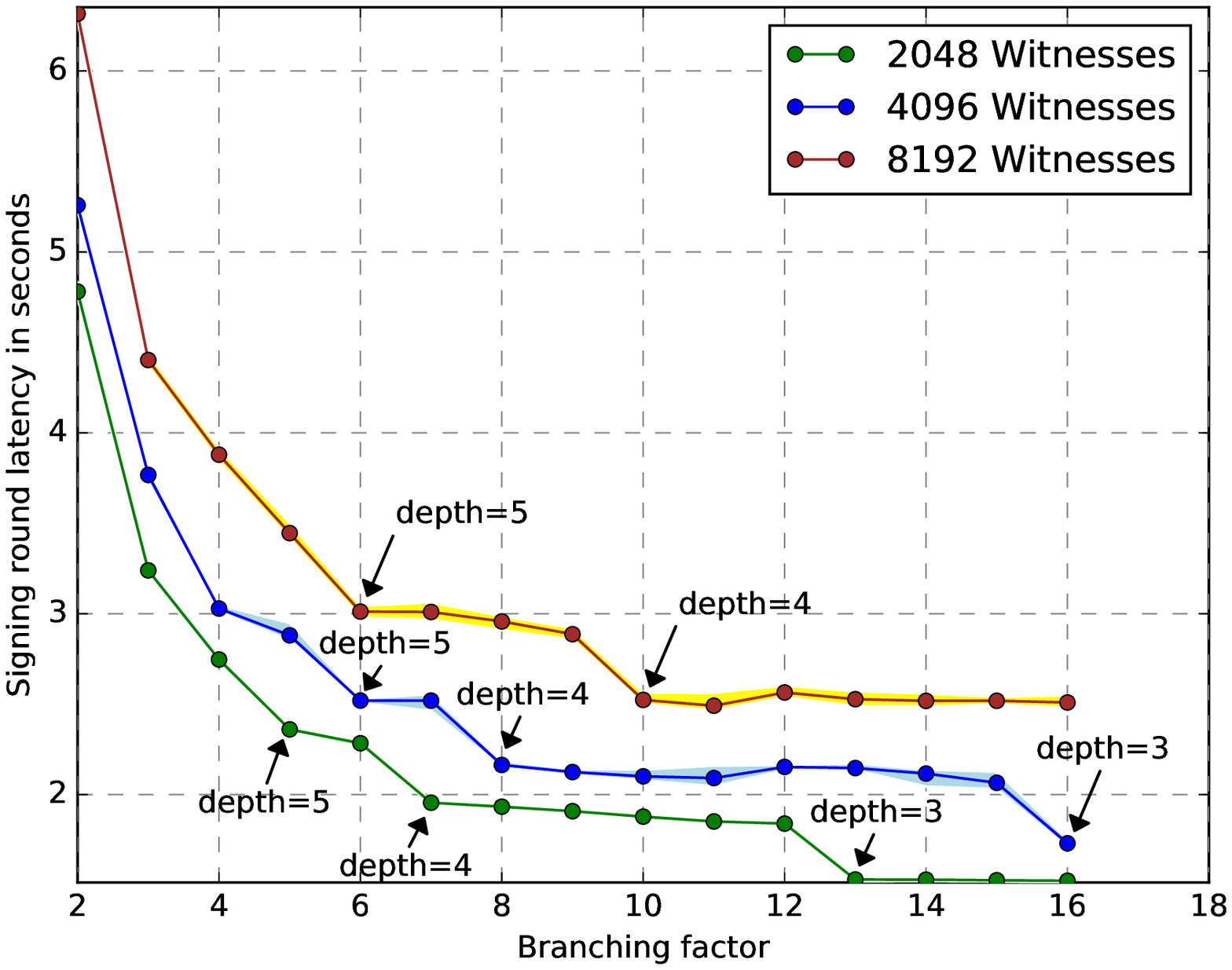}}
    \caption{Collective signing latency versus branching factor.}
\label{fig:latency-vs-bf}
\end{figure}

Our next experiment explores the tradeoffs in organizing
the spanning tree with which \cosi aggregates signatures:
in particular the tradeoffs between wide, shallow trees
and narrower, deeper trees.
This experiment is parameterized by the tree's {\em branching factor},
or maximum number of children per interior node,
where 2 represents a binary tree.

Figure~\ref{fig:latency-vs-bf} shows the relationship
between per-round signing latency and branching factor
in spanning trees containing 2,048, 4,096, and 8,192 witnesses total,
respectively.
Low branching factors increase tree depth,
increasing root to leaf round-trip latency
by about 200 milliseconds per unit of depth added.
On the other hand, low branching factors
also decrease both the CPU time spent per node
and the communication costs each node incurs coordinating with its children.

Empirically, we find that the higher the branching factor the lower the signing
latency. For example, in the case of 2,048 witnesses and a branching factor of
16, we get a tree depth of 3 and a collective signing latency of below 2 seconds.
For trees of depth 3 or less we find that computation time dominates,
while for depths 5 or more network latencies begin to dominate.
The current \cosi prototype makes no attempt to optimize its computations,
however;
further optimization of the computations
might make small depths more attractive.

\subsection{Effects of Testbed Oversubscription}
\label{sec:eval:oversub}

Since we did not have thousands of dedicated physical hosts
on which to evaluate \cosi,
we had to ``oversubscribe'' the testbed
by running multiple \cosi witness processes on each physical testbed machine.
The spanning trees are laid out such that
no two adjacent nodes in the tree run on the same physical host,
ensuring that the 200ms round-trip delays imposed by DeterLab
apply to all pairs of communicating witnesses in the tree.
However, oversubscription can introduce experimentation artifacts
resulting from compute load on each physical machine
and different \cosi witness processes' contention for other system resources;
we would like to measure the potential severity of these effects.

\begin{figure}[t]
\includegraphics[width=0.45\textwidth]{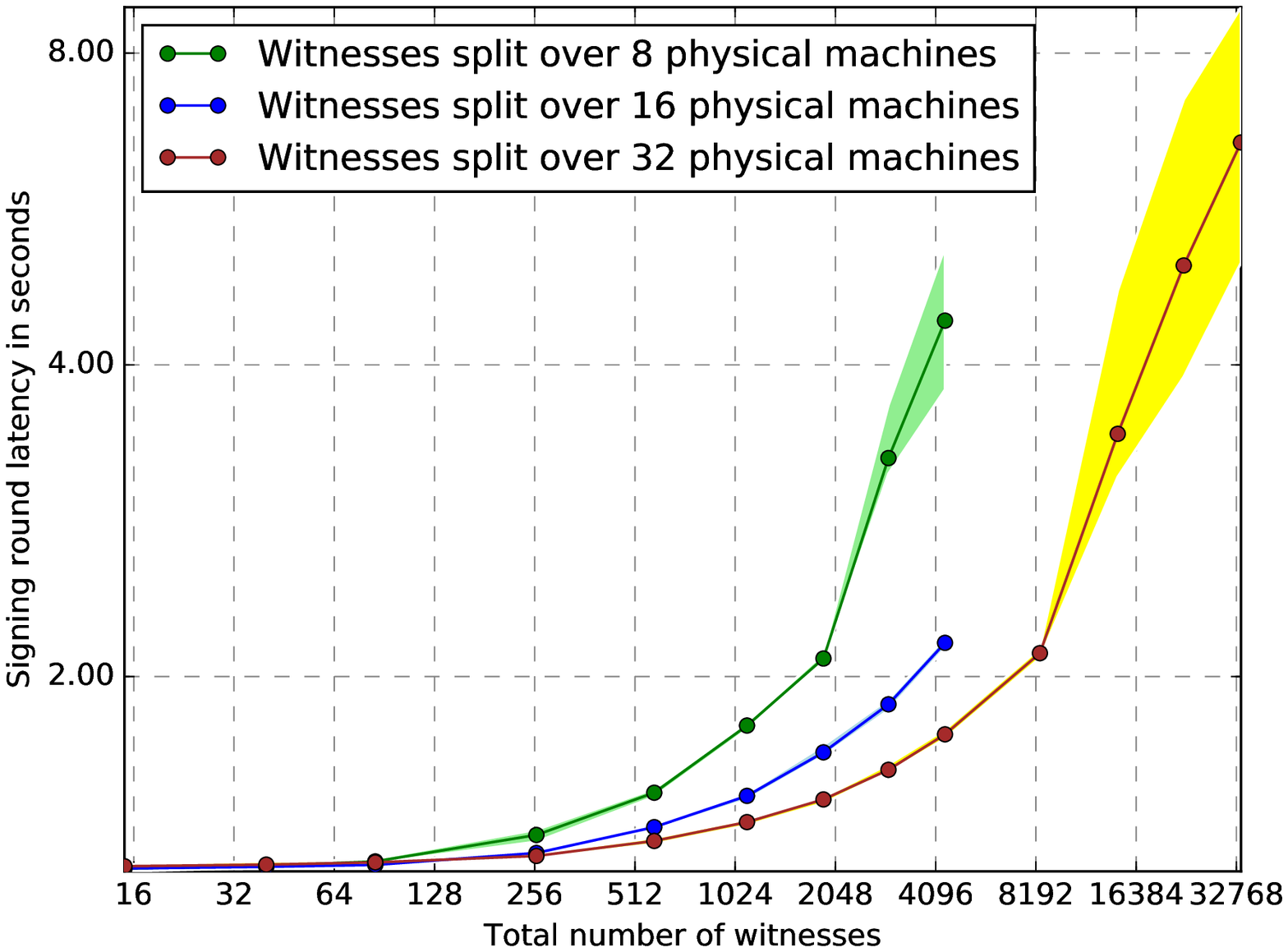}
\caption{Collective signing latency versus testbed oversubscription ratio
	(witness processes per physical machine) for a tree depth of 3.}
\label{fig:over}
\end{figure}

Figure~\ref{fig:over} shows the signing round latencies we measured
for experiments using a given number of witnesses on the $x$-axis,
but with these witness processes spread across
8, 16, or 32 physical machines
to compare different levels of oversubscription.
Unsurprisingly, the latencies become noticeably worse
at higher levels of oversubscription (fewer physical machines),
and this effect naturally increases as total number of witnesses
and hence total load per machine increases.
Nevertheless, even with these oversubscription effects
the measured latencies remain ``in the same ballpark''
for groups up to 4,096 witnesses
(512$\times$ oversubscription on 8 machines).
The performance decrease observable in
Figure~\ref{fig:latency-vs-hosts} for more than 8,192 CoSi-witnesses can be also
attributed to oversubscription and thus to the increased computational load the
32 physical machines have to handle.
Thus, since experimental oversubscription works against
\cosi's performance and scalability,
we can treat these experimental results as conservative bounds
on signing time per round;
a deployed witness cothority using dedicated
(or at least less-overloaded) witness servers may well perform
significantly better than in our experiments.

\subsection{Timestamping Application Scalability}

As discussed in Section~\ref{sec:impl:time},
our timestamping application
uses \cosi periodically to sign timestamp records
that can aggregate many clients' timestamp requests each round.
In addition, further leveraging \cosi's scalable structure,
the timestamp service allows not only the leader but also the witness servers
to handle timestamp requests from clients,
each server forming a local Merkle tree of timestamps per round
and then aggregating these local trees into one global tree
during the Commit phase of the \cosi protocol.

To evaluate the scalability of this timestamping service,
as opposed to the ``bare'' performance of \cosi signing,
we ran an experiment in which
for each \cosi server a separate process on the same physical machine
acted as a client sending timestamp requests at a constant rate.
We tested the system under a variety of client load rates,
from one request every 5 seconds to one request every 13ms --
the last case amounting to 80 requests per second on each timestamp server.
Client loads within this range did not significantly affect
the collective signing latencies we observed, however,
so we omit these graphs.

At large-scale experiments with 4,096 timestamp/witness servers
spread across 16 physical testbed machines (256 servers per machine),
each physical machine effectively handled an aggregate client load
of about 20,000 timestamp requests per second,
or 320,000 timestamp requests per second across the 4096-server collective.
Further, the current \cosi implementation and timestamp server code
is largely unoptimized and completely unparallelized within each server:
with more powerful, unshared machines,
we expect that each server could readily handle
much larger timestamping service loads.

\subsection{Difficulty of Retrofitting Existing Authorities}

Finally, to provide an informal sense for the software implementation costs
of retrofitting existing authority systems to support witness cosigning,
we relate our experience adapting the CT log server.
In this case, the log server is written in a different language (C++),
and we did not attempt to combine the log server and \cosi implementation
in a single program.
Instead, when our modified CT log server
is configured to attach collective signatures to its Signed Tree Heads (STHs),
the log server first prepares the STH internally,
then uses inter-process communication to request that
a separate process implementing the \cosi leader initiate a signing round.
The CT log server's STH signing process then waits
for the \cosi round to complete,
and incorporates the \cosi-generated collective signature
into an extension field in the STH. The verification is done in a separate
program that requests the STH from the log server and verifies the
signature against the aggregate public key of the \cosi-tree.

With this two-process approach to integrating \cosi,
the necessary changes to the CT log server
amounted to only about 315 lines as counted by CLOC~\cite{cloc},
or 385 ``raw'' source code lines.
Further, this integration took less than one person-week of effort.
While a production deployment would of course involve
significantly more effort than merely writing the code,
nevertheless our experience suggests that
it may be quite practical to strengthen existing authorities
by retrofitting them to add witness cosigning support.


\section{Discussion and Future Work}
\label{sec:disc}

This paper's primary technical focus has been on the basic \cosi protocol
for collective witnessing and signing;
we make no pretense to have addressed all the important issues
relevant to applying \cosi in any particular \coty application context.
However, we briefly revisit some of the motivating applications
introduced in Section~\ref{sec:bg}
in light of the above implementation and evaluation results.

\paragraph{Logging and Timestamping Authorities}

While the current \cosi prototype is basic,
it nevertheless already implements the essential functionality
of classic tamper-evident logging
and timestamping authorities~\cite{haber91how,adams01internet,sirer13introducing}.
As neither the leader nor any signer can produce a collective signature
without the participation of a quorum of the potentially large collective,
such a timestamp \coty can offer much stronger protection
against the equivocation, history-rewriting, or log-entry back-dating attacks
that a centralized timestamp service can mount if compromised.
When integrated into a directory~\cite{melara15coniks}
or software update service~\cite{samuel10survivable},
this timestamping architecture can offer strong proofs of freshness,
by enabling clients to submit random challenges and verify
that their challenges are included in the service's next signed update.

\com{	not sure we need this detail, it's pretty standard design...
The collective signing protocol easily extends to a secure time stamping service.
A client who wishes to timestamp a document, submits a hash to the document 
to one of the children nodes. Multiple documents can be attested to at the same time. 
The proposer announces a new round, the signatories collect and commit the data to be 
timestamped (included in the log), and produce a collective signature (add the committed 
data as a log entry).

To verify that a document was indeed included, a client uses the documents hash, 
the collective public key, and the path from the hashed document to the root with all the 
sibling node hashes to first verify that the document was truly included and then that the 
signature is valid. 
}

\paragraph{Certificate Authorities}

Adding proactive transparency and protecting clients
against stolen CA-related keys (including CT log server keys)
may be the most compelling and immediately urgent use-case for \cosi.
While adding witness cosigning to CT's log server as we explored above
represents one fairly simple and potentially worthwhile step,
more substantial modifications to the current CA system may be needed
to address other major issues
such as certificate freshness and revocation~\cite{liu15end}.

We envision that
in a witness \coty architecture in which not just one CA
but many of them inspect and collectively sign certificates,
stolen CA keys such as those of
DigiNotar~\cite{bright11diginotar,arthur11diginotar} and
Comodo~\cite{bright11comodo}
would not by themselves be usable to sign certificates
that a web browser would accept.
Not just CAs but browser vendors and security companies
could incorporate monitoring servers into the certificate cothority as signers,
to watch for and perhaps proactively impose a temporary ``veto''
on the signing of unauthorized certificates,
such as certificates proposed by a CA that is not recorded as having
contractual authority over a given domain.
Giving other CAs serving as witnesses
even temporary veto power over a CA's certificate issuance processes
creates DoS concerns,
but such concerns might be alleviated provided
administrative communication channels between CAs and witnesses are effective.

Deploying a more general certificate \coty 
would of course require addressing many additional issues
beyond the basic collective signing mechanism covered here,
not just technical but also organizational and political.
\com{
One important question is whether the computation or bandwidth costs
of allowing every CA to cross-check every other CA's certificates
might be prohibitive.
We were unable to find good statistics on the total aggregate rate
of new certificate production across all of today's CAs,
and have not implemented full certificate cross-checking
in our current \cosi prototype.
...
}
One important technical challenge is backward compatibility
and incremental deployment.
We anticipate that current root CAs might gradually
transition their root signing keys into witness cothority keys,
with their current sets of delegated CAs
(and any other cooperating root CAs) serving as witnesses.
Each root CA could transition independently at its own pace,
driven by pressure from users and browser vendors to increase security.
Web browsers would need to be upgraded gradually to support
aggregation-compatible signature schemes such as Schnorr
in addition to the currently common RSA, DSA, and ECDSA schemes.
During their transition period root CAs
could retain traditional root CA certificates
for use in older web browsers while embedding
root \coty certificates instead into suitably upgraded browsers.
However, we leave to future work a detailed exploration and analysis
of the ``right'' way to integrate witness cosigning into the CA system.

\com{

\xxx{Not implemented by I guess worth describing?}
All CAs form a single Certificate Cothority (CC).
Any certificate must be collectively signed by the whole CC,
and not just individually signed by a single CA,
before web browsers will accept the certificate.

Each \coty signer doubles for a CT server (and perhaps a CA).
In each round, a signatory submits a root of a ever-growing Merkle tree of newly produced certificates.
Each certificate receives a single collective SCT (signed certificate timestamp)
Can add Cothority SCT using X.509v3 extension, just like CT.

\subsubsection{E-Voting Service}
Each signer casts a vote by propagating a blob containing the vote up the tree;
all votes get logged and everyone can verify them. 

}

\paragraph{Public Randomness Authorities}

While not our present focus,
the current \cosi prototype also effectively implements
a simple collective public randomness service
that could improve the trustworthiness of
public randomness authorities~\cite{beacon,random-org}.
Notice that in phase 2 of the signing protocol (Section~\ref{sec:tree-sign})
each server $i$ commits to a fresh random secret $v_i$,
contributing to a collective random secret $\sum_i v_i$
that no participant will know unless {\em all} signers are compromised
or the discrete-log hardness assumption fails.
The final response produced in phase 4 depends unpredictably and 1-to-1
on this random secret and the collective challenge $c$.
Thus, we can use the final aggregate response $\hat{r}_0$
as a per-round public random value that was collectively committed in phase 2
but will be unpredictable and uncontrollable by any participant
unless all signers are colluding.

While these random outputs will be unpredictable and uncontrollable,
our current prototype cannot guarantee that they are fully {\em unbiased},
due to its reliance on the signing exception mechanism for availability.
In particular, if a malicious leader colludes with $f$ other signers,
then the leader can control whether these colluders appear online or offline
to produce up to $2^f$ different possible final aggregate responses
with different exception-sets,
and choose the one whose response is ``most advantageous'' to the leader,
just before completing phase 4 of the protocol.
Alternative approaches to handling witness failures,
through the judicious use of verifiable secret sharing (VSS)
techniques for example~\cite{feldman87practical,stadler96publicly},
might be able to address this bias issue,
by ensuring that
{\em every} node's secret is unconditionally incorporated in the final response,
unless a catastrophic failure makes some server's secret unrecoverable
even via secret-sharing.

With these changes,
a future version of \cosi might be able to offer bias-resistant randomness
in a conventional but scalable threshold-security model,
contrasting with more exotic approaches recently proposed
using new cryptographic primitives
and hardness assumptions~\cite{lenstra15random}
or the Bitcoin blockchain~\cite{bonneau15bitcoin} for example.
We again leave exploration of this opportunity to future work.

\paragraph{Other Types of Authorities}

Integrating witness cosigning into blockchain systems
such as Bitcoin~\cite{nakamoto08bitcoin}
present interesting opportunities to improve
blockchain security and performance~\cite{kokoris16enhancing}.
The tree-based scaling techniques explored here
may also be applicable to decentralizing other cryptographic primitives
such as public-key encryption/decryption.
A large-scale cothority might collectively decrypt
ElGamal~\cite{elgamal85public} ciphertexts
at particular future dates or on other checkable conditions,
to implement time-lock vaults~\cite{rivest98timelock,mont03hp},
key escrows~\cite{denning96key}, 
or fair-exchange protocols~\cite{franklin97fair}.


\com{ discuss:
Software signing authorities, \ie, app stores?
Fair exchange authorities?
Time-lock encryption authorities?
Forward-secure E-mail?
}

\section{Related Work}
\label{sec:rel}

The theoretical foundations for \cosi and witness \coties
already exist in the form of 
threshold signatures~\cite{shoup00practical,boldyreva02threshold},
aggregate signatures~\cite{boneh03survey,lu06sequential,ma09new}, 
and multisignatures~\cite{micali01accountable,bellare06multi}.
Threshold signatures allow some subset of authorized signers to produce
a signature, however, often making it impossible for the verifier to find out 
which signers were actually involved.  
In aggregate signatures, a generalization of multisignatures, 
signers produce a short signature by combining their signatures on individual statements 
through an often serial process. 
On the other hand, multisignatures closely fit the requirements of
\app for security, efficiency and the simplicity of generation across many signers. 
However, to our knowledge these primitives have been deployed only in small groups
(\eg, $\approx 10$ nodes) in practice,
and we are aware of no prior work experimentally evaluating
the practical scalability of threshold crypto or multisignature schemes.

Merkle signatures~\cite{merkle79secrecy,merkle89certified,buchmann07merkle}
employ Merkle trees for a different purpose,
enabling a single signer to produce multiple one-time 
signatures verifiable under the same public key.

Online timestamping services~\cite{haber91how,adams01internet}
and notaries~\cite{sirer13introducing} 
enable clients to prove the existence of some data 
(\eg, contracts, research results, copyrightable work) before a certain point in time
by including it in a timestamped log entry.
Typically, a trusted third party acts as 
a timestamping authority~\cite{digistamp,freeTSA,safeTSA}
and has a unilateral power to include,
exclude or change the log of timestamped data.

\com{ Compare timestamping via Cosi with timestamping via Bitcoin.u

Good question; we didn’t really try to address that alternative since timestamping is just one of several interesting applications of CoSi but I’ll make sure we at least add some discussion to the paper.  Time stamping via Bitcoin works fine, but (a) because of Bitcoin’s slowness and probabilistic “eventual consistency” model, it requires the client to wait at least 10 minutes, and more like 50 minutes to be “really certain”, that its timestamp has “definitely” been embedded in the blockchain; (b) will likely soon require the client to pay a non-trivial amount for every timestamp, as pressure on the currently/still 1MB transactions-per-block limit increases and miners gradually stop accepting no-fee or low-fee transactions; and (c) is just a fundamentally architecturally unscalable way to deal with timestamping, since it requires every timestamp submitted by any client to be included in the blockchain data seen and validated by every Bitcoin miner or full node everywhere in the world, whereas CoSi only requires every node to know about the root and the specific Merkle paths/subtrees relevant to them.

}

Many distributed systems rely on tamper-evident
logging~\cite{li04sundr,crosby09efficient}. 
Logging services are vulnerable to 
equivocation, however, where a malicious server
rewrites history or
presents different ``views of history'' to different clients.
Solutions include weakening consistency guarantees
as in SUNDR~\cite{li04sundr}, adding trusted hardware
as in TrInc~\cite{levin09trinc} or 
relying on a trusted party~\cite{schneier99secure}.
Certificate Transparency or CT~\cite{rfc6962, laurie14CT}
and NIST's Randomness Beacon~\cite{beacon} are examples 
of application-specific logging services that exemplify 
issues related to a trusted-party design paradigm. 

Directory services such as Namecoin~\cite{durham11namecoin},
and Keybase~\cite{coyne14keybase}
use blockchains such as Bitcoin~\cite{nakamoto08bitcoin}
as a decentralized timestamping authority~\cite{kirk13could}.
With this approach,
history rewriting or equivocation attacks become difficult
once a transaction is deeply embedded in the blockchain --
but clients unfortunately have no efficient decentralized way
to {\em verify} that a timestamp transaction is in the blockchain,
other than by downloading and tracking the blockchain themselves
or by trusting the say-so of centralized ``full nodes.''
Blockchains with collectively signed transactions~\cite{kokoris16enhancing}
might address this verification weakness in the blockchain approach.

There are many proposals
to address PKI weaknesses~\cite{clark2013sok}.
Browsers such as Chrome and Firefox hard-code or
{\em pin} public keys for particular sites
such as \verb|google.com|~\cite{rfc7469,kranch15upgrading}
or particular CAs for each site --
but browsers cannot ship with hard-coded certificates or CAs 
for each domain for the whole Web.
Alternatively, browsers pin the first certificate a client
sees~\cite{soghoian11certified}
protecting a site's regular users but not new users.
TACK~\cite{marlinspike13trust}, another approach to pinning, 
offers site owners the ability to authorize TLS keys for their domain 
using a long-term TACK key they control. 
Since the client's browser must witness a pin on two different occasions, 
TACK protects users from opportunistic attackers but it does not prevent an attacker with a long-term
access to the victim's network from tricking him to accept incorrect pins. 

More recent mitigations for CA weaknesses rely on 
logging and monitoring certificates as proposed in systems like 
AKI~\cite{kim13accountable}, ARPKI~\cite{basin14arpki}, 
PoliCert~\cite{szalachowski14policert},
and CT~\cite{rfc6962,laurie14CT}.
Now deployed in the Chrome browser,
CT requires CAs to insert newly-signed certificates into public logs,
which independent auditors and monitors 
check for consistency and invalid certificates.
Even with CT,
an attacker can unfortunately still create
a fake EV certificate that the Chrome browser will accept
by stealing the secret keys of,
or secretly coercing signatures from,
only three servers: any single CA
and any two CT log servers~\cite{laurie15improving}.
If the attacker also blocks the targeted device
from gossiping with public CT servers after accepting this fake certificate,
the attacker can hide this attack indefinitely~\cite{ford16apple}.
CT's reliance on clients being able to gossip with monitors and auditors
also raises latency and privacy concerns. 

COCA~\cite{zhou02coca}
distributes the operation of a CA across multiple servers,
and Secure Distributed DNS~\cite{cachin04secure}
similarly distributes a DNSSEC~\cite{rfc4033} name service.
These systems represent precedents for \cosi's collective witnessing approach,
but distribute trust across only a small group:
at most four servers in COCA's experiments
and seven in Secure Distributed DNS.
Some of these trust-splitting protocols have used threshold signatures
as a primitive~\cite{cachin00random,cachin01secure,ramasamy06parsimonious},
as \cosi does.

The NIST Randomness Beacon~\cite{beacon} logs random values 
it produces by signing them using its own secret key and chaining them with
previously produced values.
While a dishonest beacon cannot selectively change 
individual entries, it could rewrite history from a chosen point and 
present different views of the history to different clients.
Additionally, there is no guarantee of freshness of the published randomness.
While the quality of the output is likely not affected if the beacon 
precomputes the randomness,  the beacon gets to see 
these values beforehand, leaving it vulnerable to insider attacks. 

TUF~\cite{samuel10survivable} and Diplomat~\cite{kuppusamy16diplomat}
address software download and update
vulnerabilities~\cite{bellissimo06secure,cappos08look,nullbyte14hack},
in a framework that supports threshold signing
by creating and checking multiple independent signatures.
Application Transparency~\cite{fahl14apptransparency} adapts
CT to software downloads and updates.
CoSi complements both TUF and Application Transparency
by greatly increasing the number of independent servers
an attacker must compromise in order to keep the compromise secret.

\section{Conclusion}
\label{sec:conc}

This paper has demonstrated how using
theoretically established and well-understood cryptographic techniques,
we can add efficient, scalable {\em witness cosigning}
to new or existing authority services.
Witness cosigning offers proactive rather than merely retroactive transparency,
by ensuring that an attacker who compromises the authority's secret keys
cannot individually sign a statement clients will accept
without also submitting that statement to many witnesses for cosigning,
creating a high probability of immediate detection.
By making authority keys relatively useless ``in secret,''
witness cosigning also reduces the value of an authority's keys
to attackers wishing to operate in secret,
disincentivizing attacks against the authority's keys in the first place.
The encouraging scalability and performance results
we have observed with our \cosi prototype
lead us to believe that large-scale witness cothorities are practical.
If this is the case,
we feel that there may be little remaining technical reason
to settle for the centralized, weakest-link security
offered by current designs for today's common types of critical authorities.
We can and should demand stronger,
more decentralized security and transparency from 
the Internet's critical authorities.

\subsection*{Acknowledgments}

We wish to thank
Tony Arcieri,
Dan Boneh,
Joe Bonneau,
Christian Cachin,
Justin Cappos,
Rachid Guerraoui,
Jean-Pierre Hubaux,
Ben Laurie,
Eran Messeri,
Linus Nordberg,
Rene Peralta,
Apostol Vassilev,
and the anonymous reviewers
for valuable feedback and discussion during this project.
We also wish to thank Stephen Schwab and
the entire DeterLab team for their tireless support
for our experiments.
This research was supported in part by the NSF under grants 
CNS-1407454 and CNS-1409599.

\bibliographystyle{abbrv}
\bibliography{net,sec,os,soc,theory}

\begin{thebibliography}{100}
\bibitem{weissman15how}
C.~G. Weissman.
\newblock How a man who worked for the lottery association may have hacked the
  system for a winning ticket.
\newblock {\em Business Insider}, Apr. 2015.
\end{thebibliography}

\end{document}